\documentclass[lettersize,journal]{IEEEtran}
\usepackage{float}
\usepackage{graphicx}
\usepackage{subfig}
\usepackage{array}
\usepackage{amsmath}
\usepackage[justification=centering]{caption}
\usepackage{textcomp}
\usepackage{url}
\usepackage{amssymb}

\usepackage{cite}
\begin{document}

\title{QoE Optimization for Semantic Self-Correcting Video Transmission in Multi-UAV Networks}

\author{Xuyang Chen,~\IEEEmembership{Student Member,~IEEE,} Chong Huang,~\IEEEmembership{Member,~IEEE,} Daquan Feng,~\IEEEmembership{Member,~IEEE,}\\Lei Luo,~\IEEEmembership{Member,~IEEE,} Yao Sun,~\IEEEmembership{Senior Member,~IEEE,} and Xiang-Gen Xia,~\IEEEmembership{Fellow,~IEEE}
\thanks{Xuyang Chen and Daquan Feng are with the College of Electronics and Information Engineering, Shenzhen University, Shenzhen 518060, China (email: chenxuyang2021@email.szu.edu.cn, fdquan@szu.edu.cn).}
\thanks{Chong Huang is with 5GIC \& 6GIC, Institute for Communication Systems, University of Surrey, Guildford, GU2 7XH, UK (e-mail: chong.huang@surrey.ac.uk).}
\thanks{Yao Sun is with the James Watt School of Engineering, University of Glasgow, Glasgow, G12 8QQ, UK (email: Yao.Sun@glasgow.ac.uk).}
\thanks{Lei Luo is with the School of Communications and Information Engineering, Chongqing University of Posts and Telecommunications, Chongqing 400065, China, and also with the School of Information and Communication Engineering, University of Electronic Science and Technology of China, Chengdu 610051, China (e-mail: luolei@cqupt.edu.cn).}
\thanks{Xiang-Gen Xia is with the Department of Electrical and Computer Engineering, University of Delaware, Newark, DE 19716, USA (e-mail: xxia@ee.udel.edu)}
}

\maketitle

\begin{abstract}
    Real-time unmanned aerial vehicle (UAV) video streaming is essential for time-sensitive applications, including remote surveillance, emergency response, and environmental monitoring. However, it faces challenges such as limited bandwidth, latency fluctuations, and high packet loss. To address these issues, we propose a novel semantic self-correcting video transmission framework with ultra-fine bitrate granularity (SSCV-G). In SSCV-G, video frames are encoded into a compact semantic codebook space, and the transmitter adaptively sends a subset of semantic indices based on bandwidth availability, enabling fine-grained bitrate control for improved bandwidth efficiency. At the receiver, a spatio-temporal vision transformer (ST-ViT) performs multi-frame joint decoding to reconstruct dropped semantic indices by modeling intra- and inter-frame dependencies. To further improve performance under dynamic network conditions, we integrate a multi-user proximal policy optimization (MUPPO) reinforcement learning scheme that jointly optimizes communication resource allocation and semantic bitrate selection to maximize user Quality of Experience (QoE). Extensive experiments demonstrate that the proposed SSCV-G significantly outperforms state-of-the-art video codecs in coding efficiency, bandwidth adaptability, and packet loss robustness. Moreover, the proposed MUPPO-based QoE optimization consistently surpasses existing benchmarks.
\end{abstract}


\section{Introduction}
\IEEEPARstart{U}{AV} networks are critical components of integrated air-space-ground communication systems, which are considered key enablers of 6G technologies \cite{P_IEEE_2020}. With decreasing costs and improved endurance, UAVs have been widely adopted in diverse applications such as geological exploration, scene modeling, forest fire monitoring, and border surveillance. Their flexibility in deployment and rapid maneuverability also make them effective for emergency communications, serving as valuable supplements to terrestrial networks \cite{UAV_survey_IOTJ}. Video backhaul is a fundamental capability for many UAV-based missions. In scenarios such as wildfire detection and border patrol, timely video feedback is essential for providing ground operators with real-time situational awareness. However, in remote or rugged environments, wireless signals are often obstructed by natural terrain features like mountains and dense forests. Additionally, the high mobility of UAVs leads to severe wireless channel fluctuations and rapid temporal variations. Moreover, limited by the transmission link budget and the power restrictions imposed by battery capacity, UAVs often experience reduced video bitrate or even stalling. 

To further reduce the bandwidth consumption of video transmission, many studies have introduced semantic communication (SemCom) techniques. Owing to its ultra-high spectral efficiency, SemCom is considered one of the key enabling technologies for future networks \cite{Sem_Survey_2023}. Unlike conventional communications that prioritize precise bit-level reconstruction, SemCom encodes information into compact semantic representations tailored to specific downstream tasks and informed by a shared semantic knowledge base. The receiver utilizes both the noisy received semantic features and the shared knowledge base to complete the intended tasks. The compact semantic representation saves a significant amount of bandwidth budget, making it highly suitable for UAV video backhaul in bandwidth-constrained environments. Nevertheless, fluctuations in network bandwidth and sporadic semantic errors can still result in playback interruptions and degraded user experience.
To better accommodate fluctuating network bandwidth and ensure the QoE of users, many studies have further incorporated adaptive bitrate (ABR) techniques. \cite{ABR_Survey_2019}. Among them, Scalable Video Coding (SVC) is a common approach in real-time streaming applications. SVC addresses bandwidth variations by structuring the video bitstream into a base layer and multiple enhancement layers. Despite its theoretical suitability for adaptive transmission, SVC exhibits three key limitations in highly dynamic network environments. First, the strict inter-layer dependency requires successful reception of the base layer for video reconstruction, making the system highly vulnerable to packet loss over unreliable channels. Second, the rigid layer structure provides only coarse-grained bitrate control, typically limited to several discrete levels, which is insufficient for scenarios demanding fine-grained adaptation. Third, encoding multiple layers simultaneously introduces significant computational overhead compared to single-layer schemes, increasing energy consumption on resource-constrained edge devices.

\subsection{Related Works}
\textit{1) UAV video transmission.} UAV wireless video transmission is primarily divided into two categories: cellular-based video transmission and UAV ad hoc network video transmission. Cellular-based UAV video transmission utilizes cellular networks to provide broadband connections for UAVs, enabling long-distance, high-quality video transmission. There is substantial existing research in this area \cite{TNASE2024,TCSVT2022,TR2019}. In \cite{TNASE2024}, the authors defined a multi-UAV-assisted cellular communication scenario, jointly optimizing UAV trajectories, UAV selection, and transmission power to maximize QoE while minimizing the total network power consumption. In \cite{TCSVT2022}, a client-server-ground \& user (C-S-G\&U) framework and a split-merge stream (SMS) algorithm for multi-link concurrent transmission were designed. \cite{TR2019} proposed two algorithms to enhance the reliability and robustness of wireless video transmission: Pixel Row Interleaving (PRI) for effective spatial error concealment and intra-frame error recovery, and Temporal Field Interleaving with Alternative Motion Compensated Prediction (TFI-AMCP) for automatic error concealment across at least four consecutive frames.

UAV ad hoc network video transmission utilizes its networking capabilities to establish temporary UAV networks, serving as an effective supplement in areas lacking cellular base stations \cite{TVT2022 ,IOTJ2024,TCOM2021,TGCN2021}. In \cite{TVT2022}, a multi-UAV wildfire video monitoring communication framework was proposed. UAV-UEs transmit real-time video streams to UAV-BSs acting as base stations, optimizing the placement of UAV-UEs and UAV-BSs, as well as video resolution and transmission power to maximize QoE. A communication framework integrating Intelligent Reflecting Surfaces (IRS) with multi-antenna UAVs was introduced in \cite{IOTJ2024}. The authors minimized UAV propulsion energy consumption and communication energy consumption through a two-stage optimization process using path discretization, alternating optimization, successive convex approximation, and a double-loop iterative algorithm based on penalized block coordinate descent. The authors of \cite{TCOM2021} and \cite{TGCN2021} both considered UAV secure communication scenarios, employing deep reinforcement learning (DRL) techniques to select optimal video compression methods and transmission power levels to ensure user QoE and reduce energy consumption.

\textit{2) Semantic video transmission.} Video data accounts for a significant portion of internet traffic \cite{MONTIERI2021108529}. Due to the ability of SemCom to significantly reduce spectrum consumption in wireless communications, video semantic transmission has become a hot research topic \cite{JSAC_DeepWiVi,JSAC_DVSC,JSAC_SVC}. In \cite{JSAC_DeepWiVi}, a deep learning-based wireless video transmission scheme integrates video compression, channel coding, and modulation into a single neural network, known as deep joint source-channel coding (DeepJSCC). This method achieves smooth degradation when mismatched with channel quality to overcome the cliff effect of traditional digital communication schemes. In \cite{JSAC_DVSC}, a deep video semantic transmission framework based on nonlinear transformation was proposed. The authors utilized temporal priors provided by the feature domain context to learn adaptive nonlinear transformation functions for a more accurate entropy model. In \cite{JSAC_SVC}, only key points of conference videos, rather than full video frames, were transmitted, saving substantial transmission resources. The authors also utilized Channel State Information (CSI) to optimally allocate key points of varying importance levels across subchannels to further enhance performance.

Beyond video semantic transmission aimed at video reconstruction, there are some works oriented toward other machine vision applications \cite{TIP_VCM}. In \cite{TPAMI_VCM}, the authors combined the advantages of traditional video codecs at pixel-level signal processing with the strengths of neural networks in extracting semantic representations, proposing a novel framework for video semantic coding. An attention-based cross-stream feature fusion and input-adaptive feature learning method was proposed. The decoded video streams are not only visually realistic but also semantically rich, performing significantly better than benchmarks in downstream tasks such as multi-object tracking and video object segmentation. The concept of a digital retina was first introduced in \cite{TCSVT_Retina}. Within the digital retina framework, video streams are utilized for human vision, while compact feature streams are used for machine vision.


\textit{3) Adaptive bitrate.} ABR technology is commonly used to address fluctuations in network bandwidth \cite{survey_ABR}. Users must select an appropriate video bitrate based on their predicted bandwidth and buffer status. ABR techniques can generally be divided into two categories: those based on Model Predictive Control (MPC) and those based on DRL. MPC-based methods can be further divided into three types: bandwidth prediction \cite{FESFESTIVE}, buffer occupancy \cite{TON_BOLA}, and hybrid/control theory \cite{SIGCOMM_control}. In \cite{FESFESTIVE}, a robust general adaptive framework for video was proposed. The authors designed a robust chunk scheduling, bandwidth estimation, and bitrate selection mechanism, using the Jain fairness index \cite{jain1984quantitative} to ensure fairness among users. A buffer occupancy-based Lyapunov algorithm (BOLA) was introduced in \cite{TON_BOLA}. The authors formalized the bitrate adaptation problem as a utility maximization problem that considers both video quality and buffering. BOLA achieves near-optimal utility without the need to predict available network bandwidth and has been integrated into the standard DASH reference player. In \cite{SIGCOMM_control}, the authors proposed a widely used QoE standard to date. This QoE includes four parts: average bitrate, rebuffering time, average bitrate variation, and startup delay. 

Beyond MPC methods, DRL-based bitrate adaptation methods have also been extensively studied \cite{SIGCOMM_PENSIEVE,MOBICOM_OnRL,JSAC_HUANG}. A video streaming system using the DRL ABR algorithm was introduced in \cite{SIGCOMM_PENSIEVE}. This method requires no pre-programmed models or assumptions, directly inputting raw observations such as throughput and buffer occupancy into the policy network to make bitrate decisions. In \cite{MOBICOM_OnRL}, an online reinforcement learning framework was developed to improve mobile video calling. The authors associated a personalized reinforcement learning model with each user and also aggregated their experiences to form an advanced model that can handle unknown network conditions. A novel meta-reinforcement learning (meta-RL) method called A2BR (Adaptive Adaptive Bitrate Algorithm) was proposed in \cite{JSAC_HUANG}. The authors modeled the video playback process as an input-driven Markov decision process (IMDP), learning meta-strategies for different network conditions to quickly adapt the meta-strategy to the specific network environment of each user.

\subsection{Motivations and Contributions}
UAV video backhaul confronts numerous challenges, including fluctuations in network bandwidth, limited transmission link budgets, and constrained battery life. While existing semantic video transmission methods can partially alleviate bandwidth pressure, they typically perform joint compression across multiple video frames without effectively handling burst semantic errors. Consequently, the loss of a key frame can lead to decoding failure of the entire Group of Pictures (GOP), causing severe video playback interruptions. Although reducing the GOP interval can decrease stalling duration, it introduces more key frames and increases bandwidth consumption. Moreover, ABR-enabled semantic video transmission methods generally support only a limited number of discrete bitrate levels, and typically require presetting. This makes them unsuitable for complex and variable UAV networks. 

To address these challenges, this paper proposes a semantic self-correcting video transmission framework with ultra-fine bitrate granularity. The proposed system supports highly granular bitrate control and leverages inter-frame semantic correlations to enable self-correction of semantic errors. Additionally, we explore the impact of downlink resource allocation on the QoE of video transmission and demonstrate the superiority of the proposed optimization algorithm through extensive evaluations. Specifically, the main contributions of this paper are summarized as follows:
\begin{itemize}
    \item A novel semantic self-correcting video transmission framework with ultra-fine bitrate granularity (SSCV-G) is proposed. Multiple UAVs capture aerial videos from different perspectives and transmit them to the ground user for real-time monitoring. To do so, UAVs equipped with video semantic encoders map video frames to a semantically indexed space and send the corresponding indices to the ground user.
    \item A multi-frame recovery algorithm based on spatio-temporal vision transformers (ST-ViT) is proposed. The ground user recovers the current playback frame based on several received video frames. The proposed algorithm enables ultra-fine-grained bitrate adaptation by selectively discarding portions of the semantic indices according to bandwidth constraints. At the same time, it achieves semantic self-correction by exploiting the correlations across multiple received video frames.
    \item A comprehensive long-term QoE metric is defined, and a multi-user PPO (MUPPO) DRL algorithm is developed to jointly optimize UAV flight trajectories, transmission power allocation, and bitrate selection within the SSCV-G framework to maximize the long-term QoE in dynamic wireless environments.
    \item Extensive experiments fully validate the superiority of the proposed semantic self-correcting and ultra-fine-grained adaptive video transmission algorithm. The proposed joint MUPPO and SSCV-G scheme achieves significant gains in coding efficiency, bandwidth utilization, and packet-loss resilience compared to state-of-the-art video transmission baselines, particularly under constrained link budgets and highly dynamic network conditions.

\end{itemize}
\section{System Model and Problem Formulation}
In this section, we establish the system model for UAVs transmitting monitoring video. We construct a QoE model based on key perceptual metrics of video and formulate the optimization problem that needs to be addressed based on this QoE model.
\begin{figure}[t]
\includegraphics[width=\linewidth]{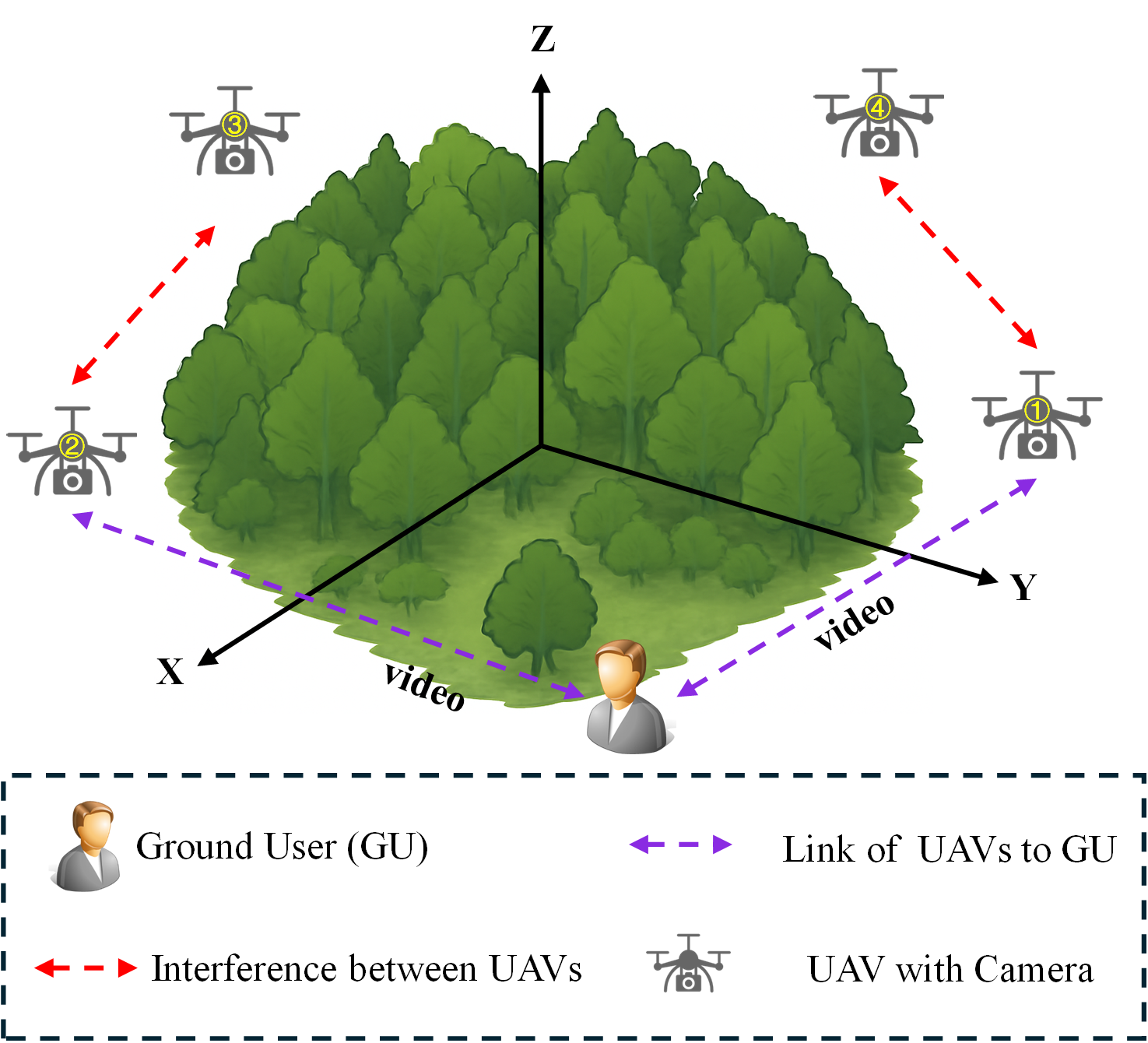}
\captionsetup{justification=raggedright}
\caption{The multi-UAV video streaming framework.}
\label{UAV_framework}
\end{figure}
\subsection{System Model}
Consider the use of $M$ UAVs, $m\in\mathcal{M}=\{1,2, ..., M\}$, for forest environment monitoring. The UAVs fly to designated airspace to capture video and transmit live streams back to the ground user. $r_0$ denotes the radius of the observed area, and the center coordinates of the area are $(x_0, y_0)$. The video stream transmission task lasts for $T$ time slots, with each slot having a duration of $\hat{t}$. The position of UAV at time $t$ is ${U_t}(x_m,y_m,z_m)$, where $(x_m,y_m,z_m)$ denotes the 3D coordinates of UAV $m$. The position of the UAV at time $t+1$ is given by 
\begin{equation}
    {U}_{t+1}(x_m,y_m,z_m)={U_t}(x_m,y_m,z_m)+\Vec{a}_m^t(x,y,z),
\end{equation}
where $\Vec{a}_m^t(x,y,z)$ is the UAV’s flying vector, determining the flying direction and distance at time $t$. $(x,y,z)$ is the moved values of the UAV position in the three coordinates from time $t$ to time $t+1$. As illustrated in Fig. \ref{UAV_framework}, without loss of generality, we uniformly deploy four UAVs to monitor the target area, and the positions of the UAVs at time $t$ can be represented as
\begin{subequations}
\label{U_m}
\begin{align}
    U_t^{m=1}=\{(x_1,y_1,z_1)|x_1=x_0, y_0+a<y_1<y_0+b,\nonumber\\z_{min}<z_1<z_{max}\},\\
    U_t^{m=2}=\{(x_2,y_2,z_2)|x_0+a<x_2<x_0+b, y_2=y_0,\nonumber\\z_{min}<z_2<z_{max}\},\\
    U_t^{m=3}=\{(x_3,y_3,z_3)|x_3=x_0, y_0-b<y_3<y_0-a,\nonumber\\z_{min}<z_3<z_{max}\},\\
    U_t^{m=4}=\{(x_4,y_4,z_4)|x_0-b<x_4<x_0-a, y_4=y_0,\nonumber\\z_{min}<z_4<z_{max}\},
\end{align}
\end{subequations}
where $a=r_0+r_s$, $b=r_0+r_s+d$. $r_s$ represents the safe distance of UAVs to the monitored area, $d$ is the movable range of the UAVs. $z_{min}$ is the minimum flying height, which prevents UAVs from colliding with mountains or trees during flight. $z_{max}$ is the maximum flying height, typically regulated by local government authorities (e.g., 120 m in UK \cite{Flight_UK}). UAVs adjust their positions to capture clear videos and adapt to changes and interference in the channel. We consider both large-scale and small-scale fading, where the small-scale fading employs a Rician channel. Both UAVs and the ground user are equipped with a single antenna. Hence, the channel coefficient between UAV $m$ and the ground user is
\begin{equation}
    h_m^t=\sqrt{\frac{\rho}{\rho+1}}\overline{H}_m^t+\sqrt{\frac{1}{\rho+1}}\hat{H}_m^t,
\end{equation}
where $\rho$ represents the Rician factor. $\overline{H}_m^t=\overline{g}_m^tG\left(\frac{c}{4\pi f_cd_m^t}\right)^{\alpha_L}$ and $\hat{H}_m^t=\hat{g}_m^tG\left(\frac{c}{4\pi f_cd_m^t}\right)^{\alpha_N}$ are the line-of-sight (LoS) and the non-line-of-sight (NLoS) components in the Rician channel, respectively. $G$ is the channel power gain factor introduced by the amplifier and antenna. $c$ denotes the speed of light. $f_c$ is the carrier frequency. $\alpha_L$ and $\alpha_N$ are the path loss exponents for LoS and NLoS in the Rician channel, respectively. $\overline{g}_m^t$ and $\hat{g}_m^t$ are the channel coefficients of LoS and NLoS, respectively. $|\overline{g}_m^t|=1$ and $\hat{g}_m^t\sim \mathcal{C}\mathcal{N}(0,1)$. $d_m$ is the distance between UAV $m$ and the ground user, which is represented as
\begin{equation}
    d_m^t=\sqrt{(x_g-x_m^t)^2+(y_g-y_m^t)^2+(h_g-z_m^t)^2}.
\end{equation}
$(x_g,y_g,h_g)$ is the position of ground user. Hence, the downlink transmission rate from UAVs to ground users at time $t$ is
\begin{equation}
    R_m^t=B\log_2\left(1+\frac{P_m^t|h_m^t|^2}{B\sigma^2+\sum_{j=1,j\neq m}^{M}P_j^t|h_j^t|^2}\right),
\end{equation}
where $B$ is the bandwidth between UAV $m$ and the ground user, $P_m^t$ denotes the transmit power for UAV $m$ in time slot $t$, $\sum_{j=1,j\neq m}^{M}P_j^t|h_j^t|^2$ is the interference from other UAVs, $\sigma^2$ is the noise power spectral density.
\subsection{QoE Model}
The QoE for real-time video streaming is typically associated with four key metrics: 1) the average bitrate of the video, 2) rebuffer time, 3) initial frame delay, and 4) smoothness of bitrate switching \cite{SIGCOMM_control}. As UAVs have identical device parameters, we disregard the capturing and encoding time of UAVs and the decoding time of the ground user. The transmission delay within a time slot $t$ is related to the chosen transmission bitrate and channel capacity, and is denoted by
\begin{equation}
    T_m^t=\frac{b_m^t}{R_m^t},
\end{equation}
where $b_m^t$ is the size of the video chunk.  The video plays normally when the transmission delay is below the delay constraint; otherwise, the video freezes. The delay threshold is related to the buffer size and caching strategy of the ground user’s player. The ground user, equipped with multi-core processors, can parallel process the videos returned by all UAVs, hence the overall transmission delay within the slot $t$ is $T^t=\max\{T_m^t\}$ \cite{JSTSP_Car_2012}. The average video bitrate is determined by the average bitrate of all video frames; a higher video bitrate provides a clearer video display but also occupies more bandwidth. The extent of bitrate changes is also closely linked to the viewer's QoE; abrupt bitrate changes can cause viewer discomfort. This paper does not consider the initial frame delay of the video. Thus, the QoE of the ground user at time slot $t$ can be represented as
\begin{equation}
\begin{aligned}
    QoE(t)=\frac{1}{M}\sum_{m=1}^M\left(\alpha q(V_m^t) - \beta|q(V_m^{t})-q(V_m^{t-1})|\right)\\-\gamma(T^t-T_c)_{+},
\end{aligned}  
\end{equation}
where $V_m^t$ is the selected bitrate, $\alpha$, $\beta$, and $\gamma$ are the weights of video quality, video smoothness, and rebuffer time, respectively. $(\cdot)_{+}=\max\{\cdot,0\}$ ensures that this term is non-negtive. $T_c$ denotes the delay constraint. $q(V_m^t)$ is the video quality metric, which can be denoted as
\begin{equation}
    q(V_m^t)=\log\left(\frac{V_m^t}{V_{\min}}\right),
\end{equation}
where $V_{\min}$ is the minimum video bitrate that can be selected.
\subsection{Problem Formulation}
\begin{figure*}[t]
    \centering
    \includegraphics[width=0.9\linewidth]{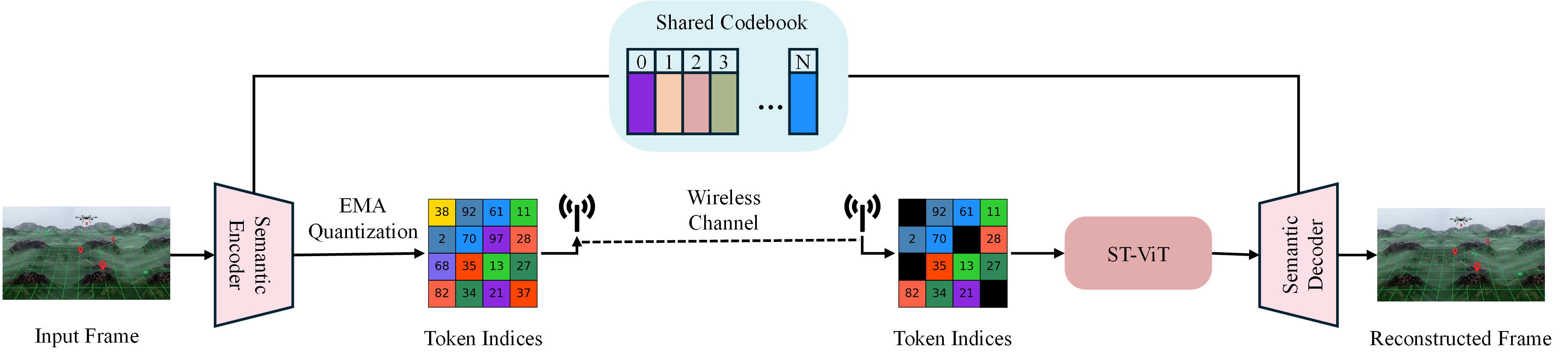}
    \captionsetup{justification=raggedright}
    \caption{The Framework of the proposed SSCV-G. ST-ViT exploits the semantic correlations among multi-frame semantic indices to reconstruct the semantic indices of the current frame.}
    \label{SSCV-G_framework}
\end{figure*}
Our objective is to find the optimal trajectory, transmit power, and video bitrate selection of UAVs to maximize the long-term average QoE. Thus, the optimization problem can be formulated as
\begin{subequations}
    \begin{align}
        \label{qoe}
        \textbf{(P0)} \quad &\underset{\{P_m^t,\vec{a}_m^t,V_m^t\}}{\max}\sum_{t=1}^TQoE(t)\\
        \label{height}
        \text{s.t.}\quad &z_{min}\leq z_m\leq z_{max}, \forall m \in \mathcal{M}.\\
        \label{power}
        &0 < P_m^t\leq P_{max},\forall m \in \mathcal{M},\\
        \label{bitrate}
        &V_{min}\leq V_m^t\leq V_{max},\forall m \in \mathcal{M},\\
        \label{position}
        &U_m^t \in Eq. (\ref{U_m}), \forall m \in \mathcal{M},\\
        \label{movement}
        &0\leq|\vec{a}_m^t|\leq a_{max}, \forall m \in \mathcal{M},
    \end{align}
\end{subequations}
where (\ref{qoe}) characterizes the QoE of all UAVs over a duration of $T$. (\ref{height}) represents the height restrictions for all UAVs. (\ref{power}) is the transmission power limit for the UAVs. (\ref{bitrate}) denotes the constraints on video quality selection. (\ref{position}) is the constraint on the placement of UAVs to ensure that there are no blind spots within the forest monitoring area. (\ref{movement}) represents the distance constraints on UAV movement within a time slot $t$, which is typically determined by the maximum flying speed of the UAVs.
\section{The proposed SSCV-G Framework}
Video streaming typically employs ABR technology to accommodate fluctuations in network bandwidth. However, conventional ABR methods typically support only a limited set of discrete bitrate levels, struggling to cope with rapidly changing wireless channels and failing to fully utilize all available bandwidth. Additionally, existing approaches that employ multi-frame joint encoding suffer from error propagation. If a key frame is lost, the corresponding video chunk must be retransmitted. To address these challenges, the SSCV-G framework is proposed, comprising four parts: 1) a video semantic encoder that encodes video frames into a series of semantic indices, 2) a bitrate controller that selectively discards semantic indices in response to real-time bandwidth constraints, 3) a packet loss recovery module that exploits temporal correlations across frames to reconstruct lost semantic indices, and 4) a video semantic decoder that maps semantic indices back to the original image. Each component is described in detail below.
\subsection{Semantic Video Codec}
The semantic video codec consists of four parts: a video semantic encoder, a video semantic decoder, a discriminator, and a codebook. To prevent error propagation caused by the loss of key frames, the semantic video encoder encodes each frame individually. The encoder includes multiple residual blocks composed of convolutional neural networks (CNN). After passing through group normalization and a sigmoid activation layer, the original image is mapped to the semantic embedding space. All vectors in the semantic embedding space are mapped to the closest codeword in the codebook. The codebook contains a discrete representation of the semantic feature space, significantly reducing the search space during generation. The semantic decoder also includes multiple residual blocks of CNN. After passing group normalization and upsampling, the encoded semantic features are mapped to the original image. An attention layer is added at the lowest resolution to aggregate contextual information. The final generated images and the original images are fed into a discriminator. The decoder strives to generate images that are as realistic as possible to deceive the discriminator, while the discriminator aims to accurately distinguish between generated and real images. In this generative adversarial setting, the decoder and discriminator converge to an optimal saddle point, at which the decoder is able to produce clear and realistic images.
\subsection{Bitrate Controller}
Video streaming often adjusts its compression levels to modify the output bitrate or offer several bitrate versions to choose from based on channel quality. However, fixed and limited bitrate versions cannot adapt to all channel bandwidths. The proposed bitrate controller can actively drop packets according to channel bandwidth, supporting more granular bitrate adjustments. Only the indices of these semantic features in the codebook need to be transmitted after encoding. On the decoder side, the corresponding semantic features are retrieved using the codebook and indices, so the transmission size is related only to the dimension of the semantic feature space and the size of the codebook. For instance, if the semantic feature dimension is $h\times w$ and the codebook size is $S$, then the bits required for transmission are $h\times w\times\log_2S$, supporting granularity adjustments by a factor of $\log_2S$. Therefore, the bitrate controller can actively discard some indices based on the current channel bandwidth until the transmission rate approximates the target bitrate.
\begin{figure}[t]
    \centering
    \includegraphics[width=\linewidth]{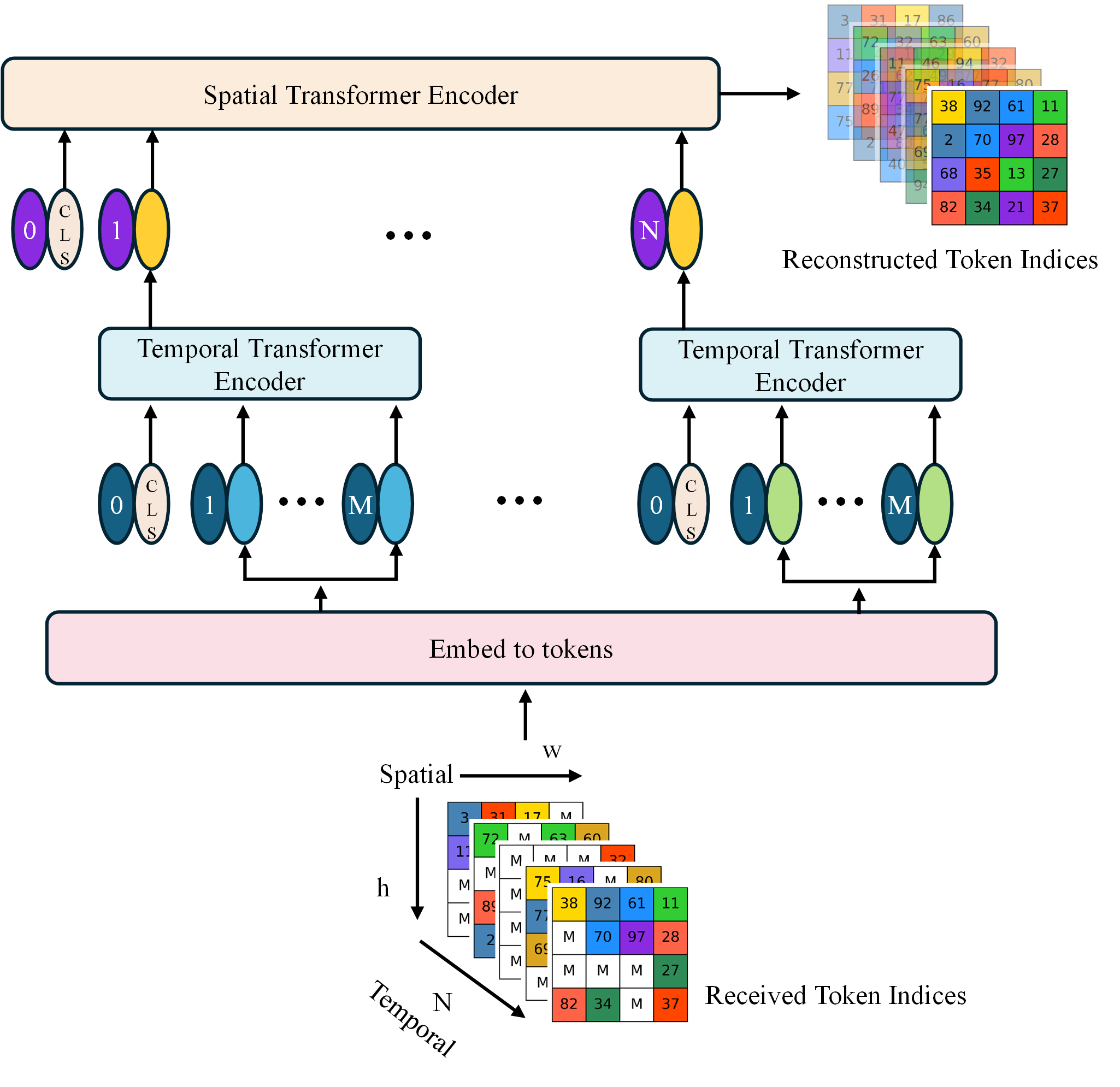}
    \caption{The proposed ST-ViT framework.}
    \label{STT}
\end{figure}
\subsection{Loss Recovery Module}
Combatting packet loss caused by bandwidth fluctuations is crucial for real-time wireless video transmission. The packet loss recovery module utilizes intra-frame and inter-frame semantic redundancy to generate the missing semantic information. For example, suppose the current frame only receives tokens corresponding to the bottom of a mountain. In that case, the peak tokens from adjacent frames can help generate the entire mountain’s tokens for the current frame. This recovery module is a neural network that takes semantic indices from the past $N$ frames as input, providing all contextual information. Learnable mask tokens are used to replace the missing indices. Transformers have become dominant in computer vision and natural language processing due to their ability to capture long-range dependencies, such as using transformers to predict missing image patches and words in \cite{He_2022_CVPR}, \cite{devlin2018bert}. Therefore, we employ a cutting-edge ST-ViT as the network structure for the loss recovery module, as shown in Fig. \ref{STT}. The ST-ViT uses attention modules at each layer to aggregate relationships between all semantic indices from past $N$ frames, weighting the correlations between received and missing indices for predicting the lost indices. Specifically, we first apply a temporal attention module across the past $N$ frames to capture inter-frame semantic correlations. The aggregated temporal features are then processed by a spatial attention module to extract intra-frame semantic dependencies. This design enables the loss recovery module to comprehensively exploit both temporal and spatial semantic correlations, thereby improving the prediction of missing tokens. It is important to note that we predict and recover based only on the semantic indices received, so decoding continues uninterrupted regardless of how many indices are lost. Moreover, the number of past frame indices used for decoding is related to the start-up delay requirements.
The training of the proposed model is divided into two stages: 1) video frame compression and reconstruction, and 2) packet loss recovery. In the first stage, the semantic video encoder $E$, the semantic video decoder $G$, the codebook $Z$, and the discriminator $D$ are jointly trained to learn compact semantic representations of video frames. In the second stage,  the parameters of $E, G, Z,$ and $D$ are frozen, and the ST-ViT module is trained to recover lost semantic indices using multi-frame semantic correlations. The training processes for these two stages are detailed below. 

\subsection{Training Semantic Video Codec}
The semantic video encoder inputs RGB video frames $\boldsymbol{x}\in \mathbb{R}^{H\times W\times 3}$ and outputs semantic features through several CNNs, with a spatial dimension of $\boldsymbol{z}\in \mathbb{R}^{h\times w\times n_z}$, i.e. $\boldsymbol{z}=E(\boldsymbol{x})\in \mathbb{R}^{h\times w\times n_z}$. The codebook $Z$ contains several discrete vectors, i.e. $Z=\{\boldsymbol{e}_k\}_{k=1}^{K},\boldsymbol{e}_k\in \mathbb{R}^{n_z}$, considered as the discrete representation of semantic features. $K$ represents the size of the codebook. We need to find the codeword in the codebook that most closely matches the current video frame's semantic features. The quantized semantic vector at position $(i,j)$ can be expressed as
\begin{equation}
    (\boldsymbol{z_q})_{ij}=Q(\boldsymbol{z})_{ij}=\left(\underset{\boldsymbol{e}_k\in Z}{\operatorname{argmin}}\|\boldsymbol{z}_{ij}-\boldsymbol{e}_k\|\right)\in \mathbb{R}^{n_z},
\end{equation}
where $Q(\cdot)$ is the quantization operation, and $(\boldsymbol{a})_{ij}$ or $\boldsymbol{a}_{ij}$ $\in \mathbb{R}^{n_z}$ is the $(i,j)$ spatial vector of $\boldsymbol{a}$. The semantic video decoder inputs the chosen semantic features $\boldsymbol{z_q}$ from the codebook and outputs the generated image $\boldsymbol{\hat{x}}$ after a series of CNNs and upsampling blocks, represented by
\begin{equation}   \boldsymbol{\hat{x}}=G(\boldsymbol{z_q})=G\left(Q\left(E\left(\boldsymbol{x}\right)\right)\right).
\end{equation}
To prevent codebook collapse caused by sparse activations in traditional vector quantization (VQ), we adopt exponential moving average (EMA) updates during codebook training. In EMA vector quantization, cluster size and embedding average are used to update the codebook entries stably. These statistics track the usage frequency of each codebook entry and the distribution of input vectors via EMA, enabling robust and adaptive codebook updates. The cluster size is defined as
\begin{equation}
    c_k^{(l)}=\gamma \cdot c_k^{(l-1)}+(1-\gamma) \sum_{ij} \mathbb{I}\left(\left(\boldsymbol{z_q}\right)_{ij}=\boldsymbol{e}_k\right),
\label{cluster_size}
\end{equation}
where $c_k^{(l)}$ denotes the cluster size of the $k$-th codebook entry at step $l$, and $\gamma$ is the EMA decay rate, which controls the proportion of historical information retained during the update process. $\mathbb{I}(\cdot)$ denotes the indicator function, which equals 1 when the specified condition is true, and 0 otherwise.
\eqref{cluster_size} represents the EMA of the frequency with which each codebook entry is selected. The embedding average is represented as
\begin{equation}
    \boldsymbol{w}_k^{(l)}=\gamma \cdot \boldsymbol{w}_k^{(l-1)}+(1-\gamma) \sum_{ij} \mathbb{I}\left(\left(\boldsymbol{z_q}\right)_{ij}=\boldsymbol{e}_k\right) \cdot \left(\boldsymbol{z_q}\right)_{ij},
\label{embedding_avg}
\end{equation}
where $\boldsymbol{w}_k^{(l)}$ denotes the embedding average of the $k$-th codebook entry at step $l$. \eqref{embedding_avg} captures the EMA of the input vectors assigned to each entry. To improve numerical stability, \eqref{cluster_size} is typically further regularized using Laplace smoothing:
\begin{equation}
    \hat{c}_k^{(l)} = \frac{c_k^{(l)}+\epsilon}{n+K\epsilon}\cdot n,
\end{equation}
where $\hat{c}_k^{(l)}$ denotes the smoothed $c_k^{(l)}$, and $n=\sum_{k=1}^Kc_k^{(l)}$, and $\epsilon$ is a very small constant, e.g., $10^{-5}$. The codebook is updated by
\begin{equation}
    \boldsymbol{e}_k^{(l)}=\frac{\boldsymbol{w}_k^{(l)}}{\hat{c}_k^{(l)}}.
\end{equation}
The loss function of training EMA-VQ is
\begin{equation}
\begin{aligned}
        \mathcal{L}_\text{VQ}(E,G,Z)=|\boldsymbol{x}-\boldsymbol{\hat{x}}|+\lambda\|\boldsymbol{z}-sg[\boldsymbol{z_q}]\|_2^2,
\end{aligned}
\end{equation}
where $\mathcal{L}_\text{rec} = |\boldsymbol{x}-\boldsymbol{\hat{x}}|$ is the reconstruction loss, $sg[\cdot]$ represents the stop-gradient operation. $\lambda$ is a weighting factor controlling the update speed of the discrete semantic features in the codebook. To further enhance video perceptual quality, a discriminator is used at the receiver end to improve the generative quality of the semantic decoder, while also aiding the codebook in learning richer semantic features. The generative adversarial loss is denoted as
\begin{equation}
    \mathcal{L}_\text{GAN}\left(\{E,G,Z\},D\right)=\log\left(D(\boldsymbol{x})\right)+\log\left(1-D(\boldsymbol{\hat{x}})\right).
\end{equation}
Consequently, the loss function for training semantic video codec is represented as
\begin{equation}
    \mathcal{L}_\text{codec}=\mathcal{L}_\text{VQ}(E,G,Z)+\eta \mathcal{L}_\text{GAN}\left(\{E,G,Z\},D\right),
\end{equation}
where the adaptive weight $\eta$ is computed by
\begin{equation}
    \eta=\frac{\nabla_{GL}|\mathcal{L}_\text{rec}|}{\nabla_{GL}|\mathcal{L}_\text{GAN}|+\delta},
\end{equation}
where $\nabla_{GL}[\cdot]$ is the gradient of its input with respect to the last layer $L$ of the semantic video decoder, and $\delta$ is added for numerical stability. 
\subsection{Training Loss Recovery Module}
The first stage of training results in a semantic video codec with compression and recovery capabilities. However, it cannot counter packet losses caused by network fluctuations. The second stage utilizes the ST-ViT at the receiver end to merge intra-frame and inter-frame semantic information to combat packet loss. Specifically, during the second stage of training, we first fix all model parameters obtained from the first stage. Then, when transmitting semantic indices, we randomly drop them with a certain probability. In each training iteration, the packet loss probability is sampled from a truncated Gaussian distribution ranging from 0 to 0.6, with a mean of 0.3 and a variance of 0.3. Dropped token indices are replaced by learnable mask tokens to ensure consistent input length for the ST-ViT. Let $\boldsymbol{s}=[s_{ijk}]_{i=1,j=1,k=1}^{h,w,N}$ denote the semantic indices of past $N$ frames received at the receiver. $i$ and $j$ denote the spatial position indices within a frame, and $k$ denotes the temporal index of the frame. $\boldsymbol{M}=[m_{ijk}]^{h,w,N}_{i=1,j=1,k=1}$ represents the corresponding binary mask. $m_{i,j,k}=1$ indicates that the semantic index at this position is discarded, otherwise, it is preserved. The training objective of the ST-ViT is to predict the missing indices of the last frame (the $N$-th frame among the past $N$ frames) based on all indices of the past $N$ frames. Note that only the missing indices of the last frame are predicted, and only the indices of the last frame are fed into the video semantic decoder. We use cross-entropy to maximize the log-likelihood function, denoted by
\begin{equation}
\label{indices-loss}
\mathcal{L}_{\text {loss}}=-\sum_{i=1}^h \sum_{j=1}^wm_{ijN} \log p\left(s_{ijN} \mid \boldsymbol{s_M}\right),
\end{equation}
where $\boldsymbol{s_M}$ represents the masked semantic indices received from the past $N$ frames that are used for the prediction. $p(s_{ijN}|\boldsymbol{s_M})$ denotes the probability density function of the semantic indices at position $(i,j)$ of the frame $N$, given $\boldsymbol{s_M}$. In \eqref{indices-loss}, the frame index $k$ is set to $N$, which means that only the semantic indices of the last frame are predicted. The ground truth is set as one-hot token indices to compute the cross-entropy.
\begin{figure}[t]
    \centering
    \includegraphics[width=\linewidth]{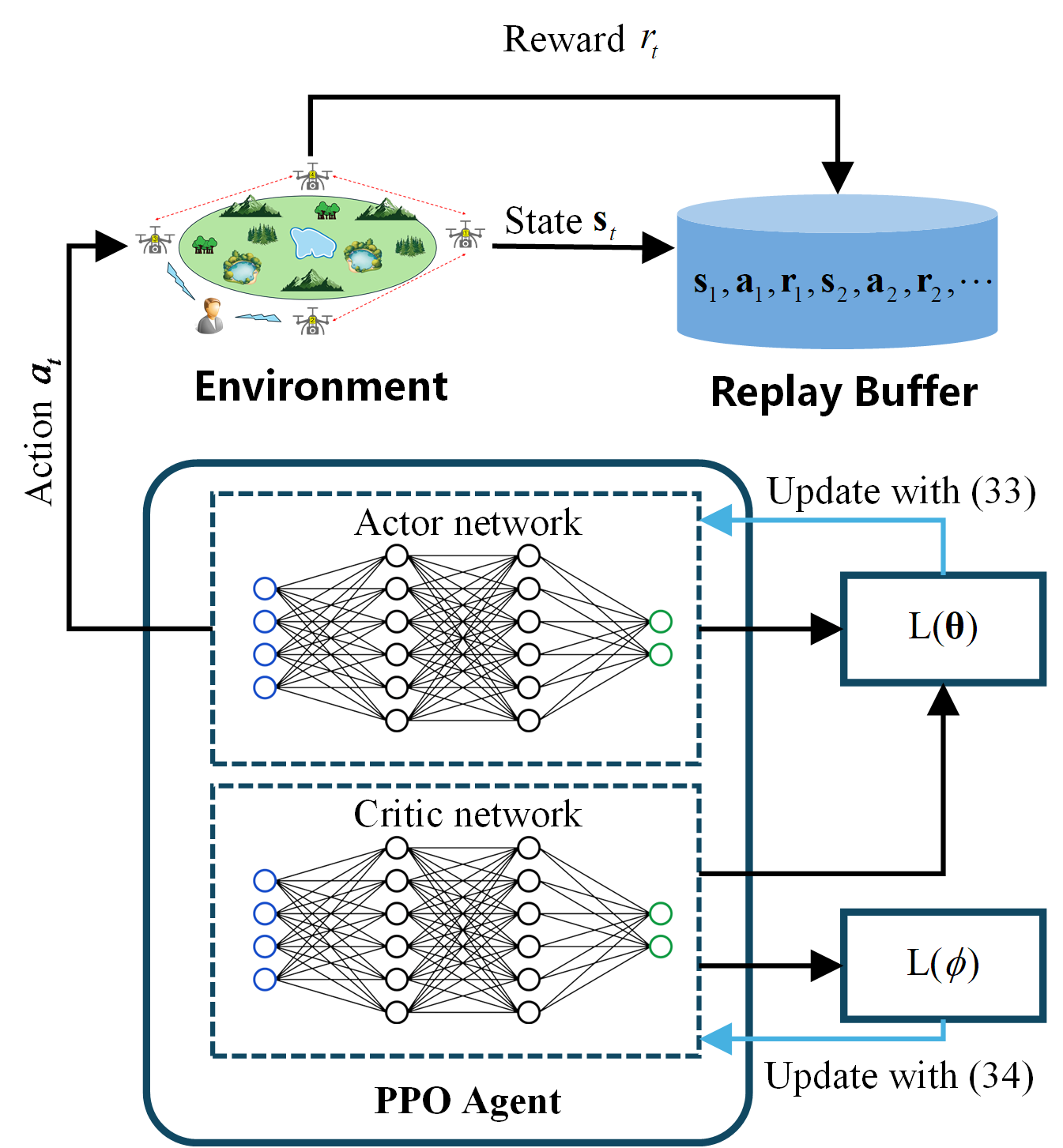}
    \caption{The framework of the proposed MUPPO.}
    \captionsetup{justification=raggedright}
    \label{MUPPO}
\end{figure}
\section{MUPPO for Resource Allocation}
To solve non-convex mixed-integer nonlinear programming (MINLP) problems \textbf{P0}, we propose the MUPPO DRL method based on action-critic mechanisms to maximize the long-term average QoE for all UAVs. We transform the non-convex optimization problem \textbf{P0} into an MDP problem and derive the training algorithm for MUPPO.

\subsection{Problem Reformulation Based on MDP}
We first define a four-tuple $\langle\mathcal{S}, \mathcal{A}, \mathcal{R}, \gamma\rangle$ to model the non-convex optimization problem as an MDP, where $\mathcal{S}$ represents the state space set, $\mathcal{A}$ denotes the action space set, $\mathcal{R}$ indicates the reward function, and $\gamma$ is the discount factor. All UAVs make decisions based on the states they observe, performing actions that yield rewards or penalties from the environment. Based on this feedback, UAVs update their strategies to maximize rewards. The state space for the UAVs at time slot $t$ is defined as $s_m(t) = \{h_m^t, P_m^t, (x_m,y_m,z_m)\}\in \mathcal{S}$. The action space is defined as $a_m(t) = \{P_m^t,V_m^t \Vec{a}_m^t(x,y,z)\}\in \mathcal{A}$. UAVs need to jointly optimize transmission power, video bitrate, and flight vectors based on the current observed state to maximize long-term QoE. Therefore, the reward function within time slot $t$ is defined as $r(t)=QoE(t)$, and the long-term cumulative reward function is defined as
\begin{equation}
    R_t(\tau) = \sum_{t^\prime=t}^T\gamma^{t^\prime-t}r_{t^\prime},
\end{equation}
where $\tau=\{s_1,a_1,s_2,a_2,\cdots,s_T,a_T\}$ is the complete episode trajectory. Let $\pi_{\boldsymbol{\theta}}$ represent the policy parameterized by $\boldsymbol{\theta}$ based on a neural network. The objective function of the policy network is then defined as 
\begin{equation}
\label{PPO_J}
    J\left(\pi_{\boldsymbol{\theta}}\right)=\mathbb{E}_{\boldsymbol{a}_t \sim \pi_{\boldsymbol{\theta}}, \boldsymbol{s}_t \sim \mathcal{P}}\left[ A\left(\boldsymbol{s}_t, \boldsymbol{a}_t\right)\right],
\end{equation}
where $A\left(\boldsymbol{s}_t, \boldsymbol{a}_t\right)$ is the advantage function for state-action pairs, which will be further elaborated upon subsequently. UAVs seek the optimal policy to maximize $J\left(\pi_{\boldsymbol{\theta}}\right)$, thus the optimization problem \textbf{P0} is transformed into
\begin{equation}
\begin{aligned}
\max _{\pi_{\boldsymbol{\theta}}} \ & J\left(\pi_{\boldsymbol{\theta}}\right) \\
\text { s.t. } \ & \boldsymbol{a}_t \sim \pi\boldsymbol{_\theta}\left(\boldsymbol{a}_t \mid \boldsymbol{s}_t\right), \boldsymbol{s}_{t+1} \sim \operatorname{Pr}\left(\boldsymbol{s}_{t+1} \mid \boldsymbol{s}_t, \boldsymbol{a}_t\right),
\end{aligned}
\end{equation}
where $\operatorname{Pr}\left(\boldsymbol{s}_{t+1} \mid \boldsymbol{s}_t, \boldsymbol{a}_t\right)$ represents the probability of taking action $\boldsymbol{a}_t$ given state $\boldsymbol{s}_t$ under policy $\pi_{\boldsymbol{\theta}}$.
\subsection{PPO-Based Resource Allocation}
Fig. \ref{MUPPO} illustrates the proposed MUPPO framework. PPO is an emerging policy gradient algorithm that employs a mini-batch gradient update method for sampled trajectories, addressing the issue of determining the optimal step size in previous policy gradient methods.  Updating with only a subset of samples helps avoid gradient oscillation, making policy updates more stable. However, mini-batch multi-step updates result in a deviation between the current policy and the policy used during interaction with the environment. Importance sampling techniques are employed to constrain this policy deviation and stabilize the learning process. By applying importance sampling to \eqref{PPO_J}, the objective function of PPO is updated to
\begin{equation}
J(\pi_{\boldsymbol{\theta}})=\mathbb{E}_{\left(\boldsymbol{s}_t, \boldsymbol{a}_t\right)} \sim \pi_{\boldsymbol{\theta}_{\text{old}}}\left[\frac{p_{\boldsymbol{\theta}}\left(\boldsymbol{a}_t \mid \boldsymbol{s}_t\right)}{p_{\boldsymbol{\theta}_\text{old}}\left(\boldsymbol{a}_t \mid \boldsymbol{s}_t\right)} A^{\boldsymbol{\theta}_\text{old}}\left(\boldsymbol{s}_t, \boldsymbol{a}_t\right)\right],
\end{equation}
where $\pi_{\boldsymbol{\theta}_{\text{old}}}$ represents the old policy that generates a trajectory through interaction with the environment, parameterized by $\boldsymbol{\theta}_{\text{old}}$. Functions $p_{\boldsymbol{\theta}}\left(\boldsymbol{a}_t \mid \boldsymbol{s}_t\right)$ and $p_{\boldsymbol{\theta}_\text{old}}\left(\boldsymbol{a}_t \mid \boldsymbol{s}_t\right)$ are the action probability densities under the policies $\boldsymbol{\theta}_{\text{old}}$ and $\boldsymbol{\theta}$ given state $\boldsymbol{s}_t$, respectively. The state-action advantage function $A^{\boldsymbol{\theta}_\text{old}}\left(\boldsymbol{s}_t, \boldsymbol{a}_t\right)$ can be represented as
\begin{equation}
\label{state-action-value}
A^{\boldsymbol{\theta}_{\text{old}}}\left(\boldsymbol{s}_t, \boldsymbol{a}_t\right)=Q^{\boldsymbol{\theta}_{\text{old}}}\left(\boldsymbol{s}_t, \boldsymbol{a}_t\right)-V^{\boldsymbol{\theta}_{\text{old}}}\left(\boldsymbol{s}_t\right).
\end{equation}
$Q^{\boldsymbol{\theta}_{\text{old}}}\left(\boldsymbol{s}_t, \boldsymbol{a}_t\right)$ represents the state-action value function, which describes the value of taking action $\boldsymbol{a}_t$ in given state $\boldsymbol{s}_t$, and is expressed as
\begin{equation}
    Q^{\boldsymbol{\theta}_{\text{old}}}\left(\boldsymbol{s}_t, \boldsymbol{a}_t\right) = \mathbb{E}_{\boldsymbol{a}_t \sim \pi_{\boldsymbol{\theta_{\text{old}}}},\boldsymbol{s}_t \sim \mathcal{P}}\left[\sum_{t^{\prime}=t}^T\gamma^{t^{\prime}-t}r_{t^{\prime}} \right].
\end{equation}
$V^{\boldsymbol{\theta}_{\text{old}}}\left(\boldsymbol{s}_t\right)$ represents the state value function, which is the expected value of the state-action value function, expressed as
\begin{equation}
    V^{\boldsymbol{\theta}_{\text{old}}}\left(\boldsymbol{s}_t \right) = \mathbb{E}_{\boldsymbol{s}_t \sim \mathcal{P}}\left[\sum_{t^{\prime}=t}^T\gamma^{t^{\prime}-t}r_{t^{\prime}} \right].
\end{equation}
$Q^{\boldsymbol{\theta}_{\text{old}}}\left(\boldsymbol{s}_t, \boldsymbol{a}_t\right)$ can be represented in another form of temporal difference (TD) as $Q^{\boldsymbol{\theta}_{\text{old}}}\left(\boldsymbol{s}_t, \boldsymbol{a}_t\right) = r_t + \gamma V^{\boldsymbol{\theta}_{\text{old}}}\left(\boldsymbol{s}_{t+1} \right)$, therefore \eqref{state-action-value} can be reformulated as
\begin{equation}
    A^{\boldsymbol{\theta}_{\text{old}}}\left(\boldsymbol{s}_t, \boldsymbol{a}_t\right)=r_t + \gamma V^{\boldsymbol{\theta}_{\text{old}}}\left(\boldsymbol{s}_{t+1} \right)-V^{\boldsymbol{\theta}_{\text{old}}}\left(\boldsymbol{s}_t \right).
\end{equation}
To better evaluate the current state-action value, a more stable and accurate measure, Generalized Advantage Estimation (GAE), is typically used to evaluate the current state-action value, expressed as
\begin{equation}
\label{GAE}
    A^\text{GAE}(\boldsymbol{s}_t, \boldsymbol{a}_t)=\sum_{l=0}^{T-t}(\lambda\gamma)^l\delta_{t+l},
\end{equation}
where $\delta_{t}=A^{\boldsymbol{\theta}_{\text{old}}}\left(\boldsymbol{s}_t, \boldsymbol{a}_t\right)$. $\lambda \in (0,1]$ is a discount factor. 
\begin{table*}[t]
\centering
\caption{Simulation Parameters}
\begin{tabular}{|>{\raggedright}p{0.3\textwidth}|p{0.1\textwidth}||p{0.32\textwidth}|p{0.12\textwidth}|}
\hline
\textbf{Training parameters for MUPPO} & \textbf{Value} & \textbf{Training parameters for semantic video codec} & \textbf{Value}\\
\hline
Number of UAVs $M$ & 4 & Optimizer & Adam\\\hline
Bandwidth B & 20 MHz & Base learning rate & $4.5\times10^{-6}$ \\\hline
Noise power spectral density $\sigma^2$ & -174 dBm/Hz & Batch size & 8 \\\hline
The radius of target region $r_0$ & 100 m & Weight decay & 0\\\hline
The flight height range & [10, 50] m & Learning rate schedule & Constant\\\hline
The flight vector & 1 m & Optimizer momentum & $\beta_1,\beta_2=0.5,0.9$\\\hline
The speed of light & $3\times10^8$ m/s & Image resolution & 240\\\hline
Transmit power & [1, 5] $W$ & \textbf{Training parameters for ST-ViT} & \textbf{Value}\\\hline
Learning rate & $5\times10^{-4}$ & Optimizer & Adam \\\hline
Advantage discount factor $\lambda$ & 0.95 & Learning rate & $1.5\times10^{-5}$\\\hline
Reward discount factor $\gamma$ & 0.99 & Weight Decay & 0.05 \\\hline
PPO-clip parameter $\epsilon$ & 0.2 & Batch size & 6\\\hline
Policy entropy bonus weight $b_2$ & 0.01 & Learning rate schedule & Cosine decay\\\hline
Time slot & 1 s & Optimizer momentum & $\beta_1,\beta_2=0.5,0.9$ \\\hline
The exponent of LoS $\alpha_L$ & 2 & Consecutive frame for loss recovery $N$ & 6 \\\hline
The exponent of NLoS $\alpha_N$ & 3 & Training epochs & 200\\\hline
\end{tabular}
\label{training_parameters}
\end{table*}
Importance sampling techniques make the probability distribution of policy $\pi_{\boldsymbol{\theta}}$ equivalent to that of policy $\pi_{\boldsymbol{\theta}_\text{old}}$ by multiplying it by an importance weight. However, the Monte Carlo sampling method used during training is not sufficient to fully characterize the probability distributions of the two policies. When the difference between the two policy distributions is significant, it can cause severe oscillations during policy updates. Therefore, regularization techniques are necessary to constrain the two distributions. Specifically, PPO employs a clipping technique to limit the discrepancy between $p_{\boldsymbol{\theta}}\left(\boldsymbol{a}_t \mid \boldsymbol{s}_t\right)$ and $p_{\boldsymbol{\theta}_\text{old}}\left(\boldsymbol{a}_t \mid \boldsymbol{s}_t\right)$, expressed as
\begin{equation}
\begin{aligned}
    &J^{\text{CLIP}}(\boldsymbol{\theta}) = \mathcal{L}_t^{\text{CLIP}}\\
    &=\mathbb{E}_t\left[\min \left(r_t(\boldsymbol{\theta})A^{\boldsymbol{\theta}_\text{old}}\left(\boldsymbol{s}_t, \boldsymbol{a}_t\right), g\left(\epsilon, A^{\boldsymbol{\theta}_\text{old}}\left(\boldsymbol{s}_t, \boldsymbol{a}_t\right)\right)\right) \right],
\end{aligned}
\end{equation}
where $r_t(\boldsymbol{\theta})=\frac{p_{\boldsymbol{\theta}}\left(\boldsymbol{a}_t \mid \boldsymbol{s}_t\right)}{p_{\boldsymbol{\theta}_\text{old}}\left(\boldsymbol{a}_t \mid \boldsymbol{s}_t\right)}$ and $g$ is the clip function which can be defined as
\begin{equation}
\begin{aligned}
    g(\epsilon, &A^{\boldsymbol{\theta}_\text{old}}\left(\boldsymbol{s}_t, \boldsymbol{a}_t\right))=\\
    &\begin{cases}(1+\epsilon) A^{\boldsymbol{\theta}_\text{old}}\left(\boldsymbol{s}_t, \boldsymbol{a}_t\right),\ \text{if} \ A^{\boldsymbol{\theta}_\text{old}}\left(\boldsymbol{s}_t, \boldsymbol{a}_t\right) \geq 0, \\ (1-\epsilon) A^{\boldsymbol{\theta}_\text{old}}\left(\boldsymbol{s}_t, \boldsymbol{a}_t\right),\ \text{if} \ A^{\boldsymbol{\theta}_\text{old}}\left(\boldsymbol{s}_t, \boldsymbol{a}_t\right)<0,\end{cases}
\end{aligned}
\end{equation}
where $\epsilon$ is a hyperparameter. When the policy ratio $r_t(\boldsymbol{\theta})$ falls within the range of $[1-\epsilon, 1+\epsilon]$, it updates normally, otherwise, it is clipped. This prevents overly rapid updates of the policy, thereby ensuring stable updates. The loss function for the critic network is defined as the error between the state value function $V^{\boldsymbol{\phi}}\left(\boldsymbol{s}_{t}\right)$ and the estimated cumulative return $R_t(\tau)$, which is denoted as
\begin{equation}
\mathcal{L}_t^{\mathrm{V}}=\left\|V^{\boldsymbol{\phi}}\left(\boldsymbol{s}_{t}\right)-R_t(\tau)\right\|^2,
\end{equation}
where $\boldsymbol{\phi}$ is the parameters of critic network. In summary, MUPPO updates the parameters $\boldsymbol{\theta}$ of actor network through
\begin{equation}
\label{update_theta}    \text{arg}\max_{\boldsymbol{\theta}}\hat{\mathbb{E}}_t\left[\mathcal{L}_t^\text{CLIP}+b_1 H\left(\pi_{\boldsymbol{\theta}}\left(\cdot\mid \boldsymbol{s}_t\right)\right)\right],
\end{equation}
where $b_1$ is a weight coefficient, and $H\left(\pi_{\boldsymbol{\theta}}\left(\cdot\mid \boldsymbol{s}_t\right)\right)$ represents the entropy reward to encourage the model to explore thoroughly. $\hat{\mathbb{E}}_t$ is the expectation approximated by using Monte Carlo sampling.
The parameters $\boldsymbol{\phi}$ of critic network are updated through
\begin{equation}
\label{update_phi}    \text{arg}\min_{\boldsymbol{\phi}}\hat{\mathbb{E}}_t\left[\mathcal{L}_t^{\mathrm{V}}\right].
\end{equation}

\begin{figure*}[t]
\centering
\subfloat[]{\includegraphics[width=0.24\textwidth]{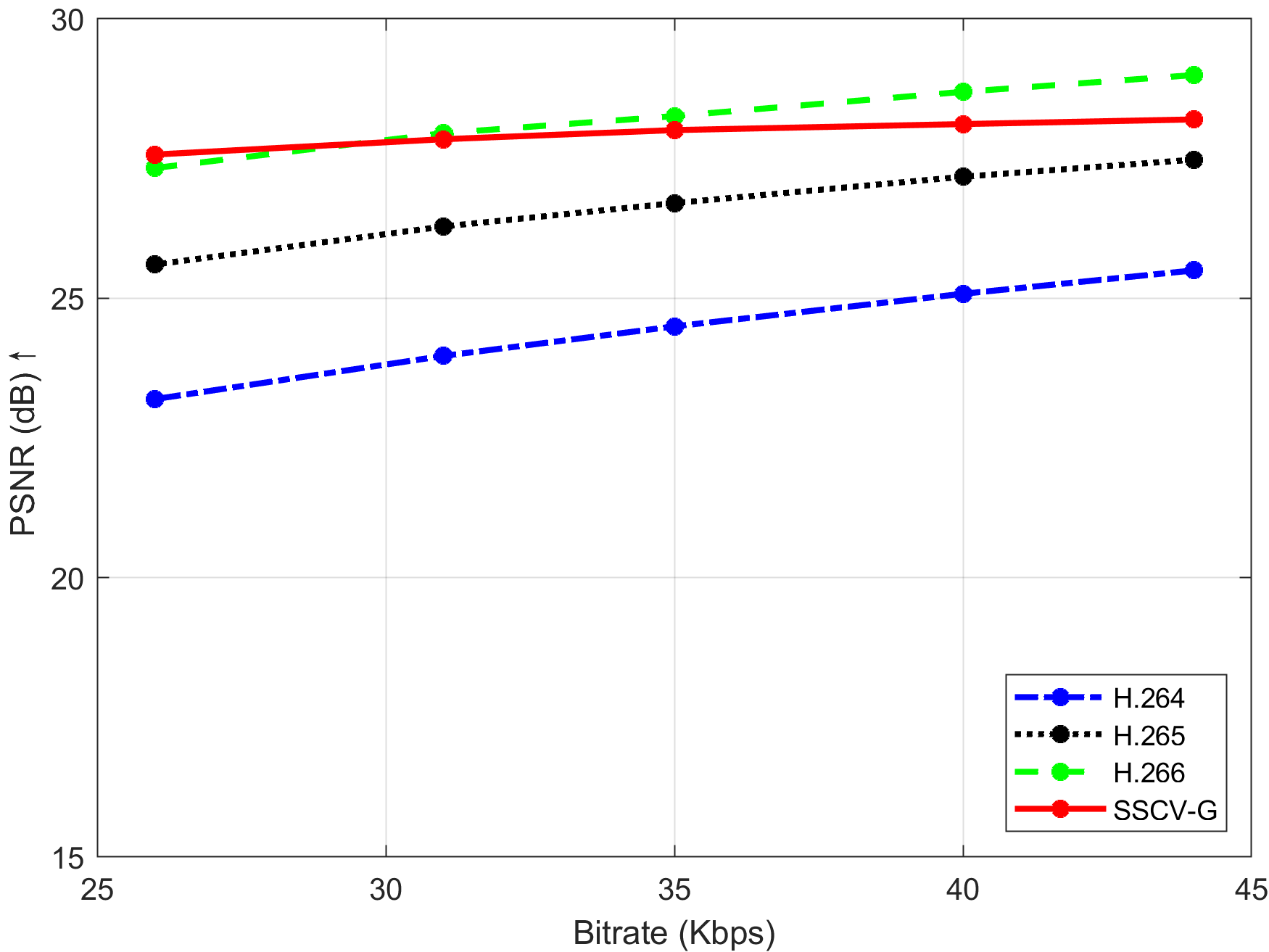}}%
\label{psnr}
\hfil
\subfloat[]{\includegraphics[width=0.24\textwidth]{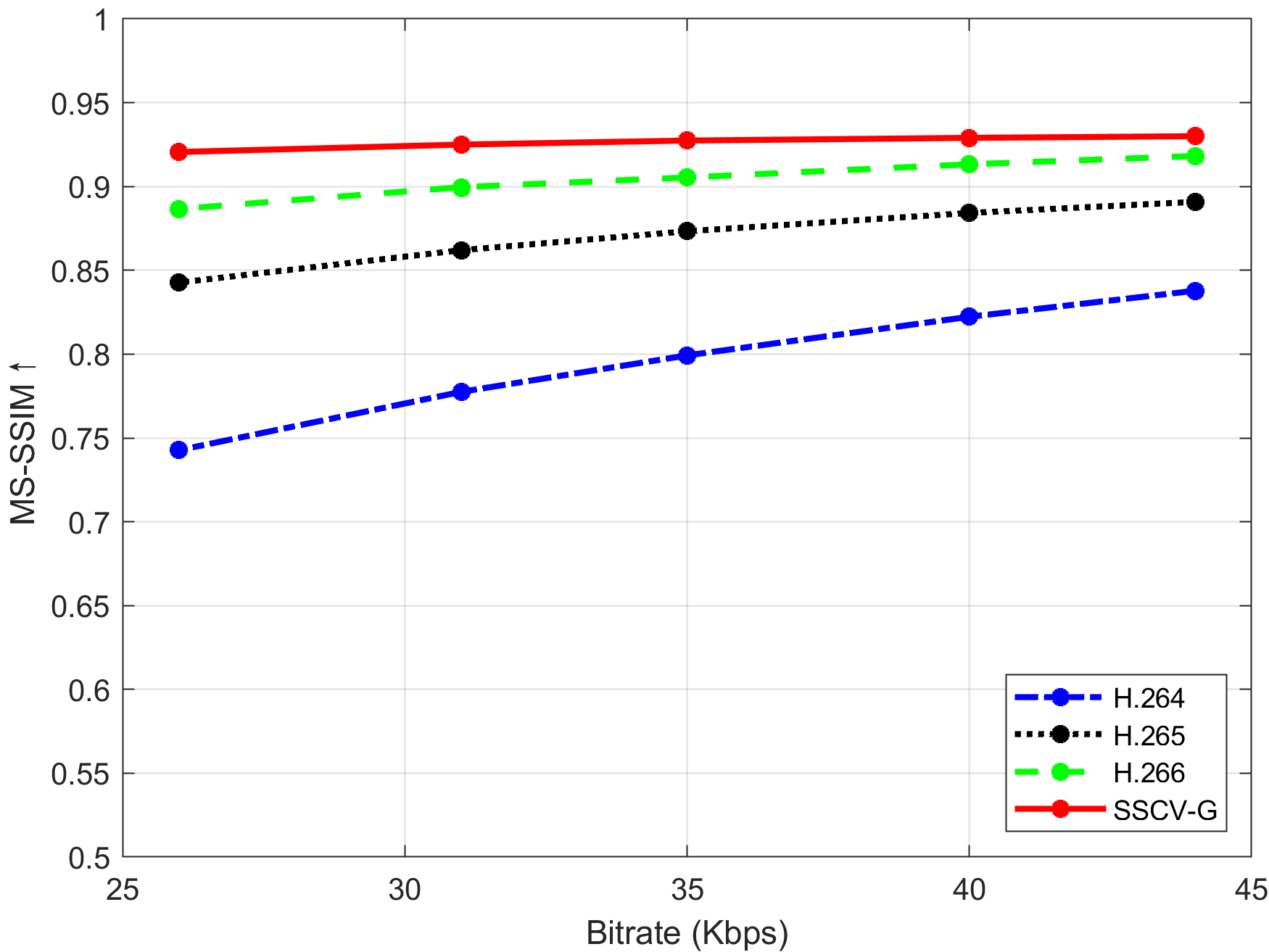}}%
\label{ms-ssim}
\hfil
\subfloat[]{\includegraphics[width=0.24\textwidth]{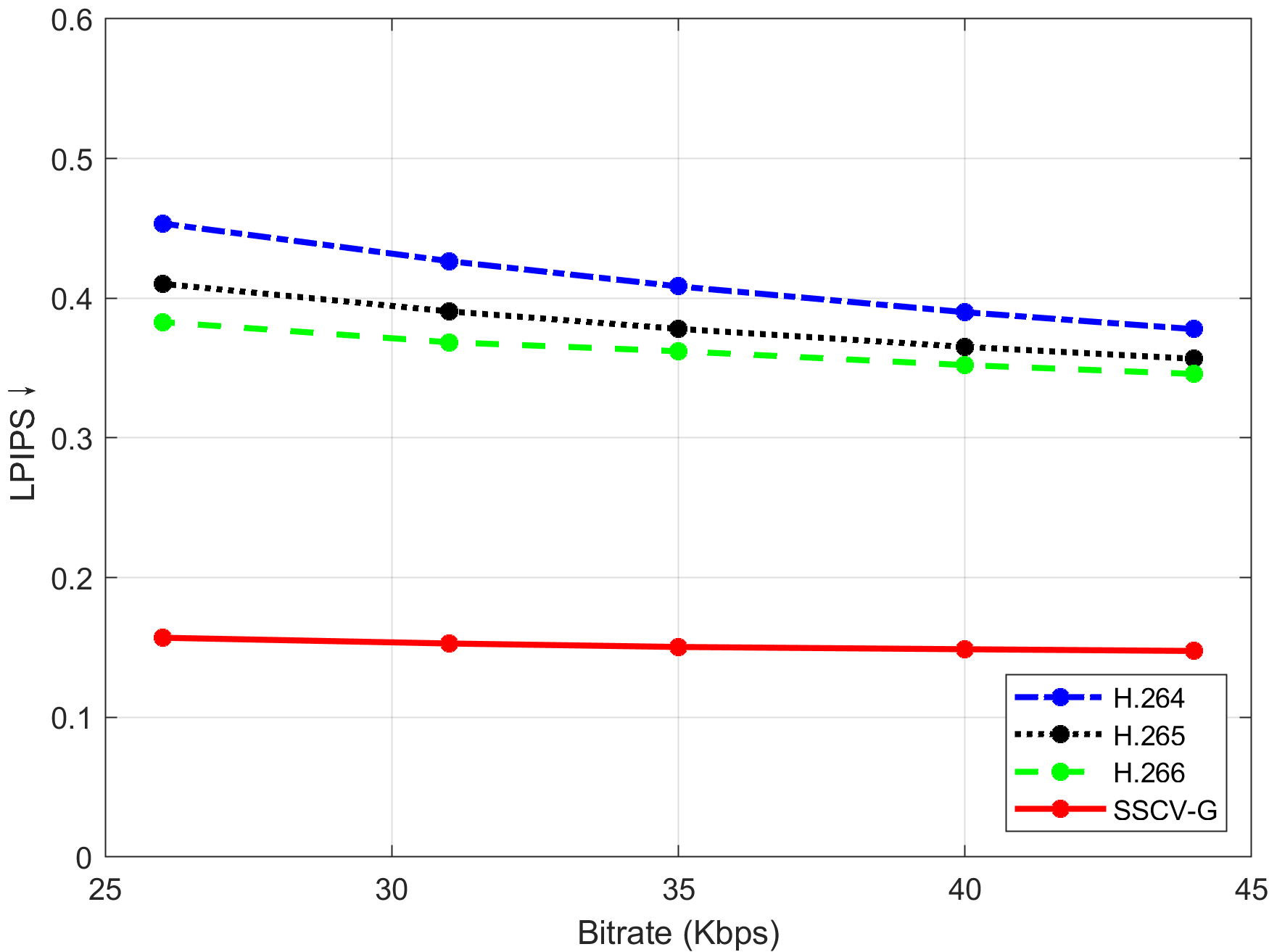}}%
\label{lpips}
\subfloat[]{\includegraphics[width=0.24\textwidth]{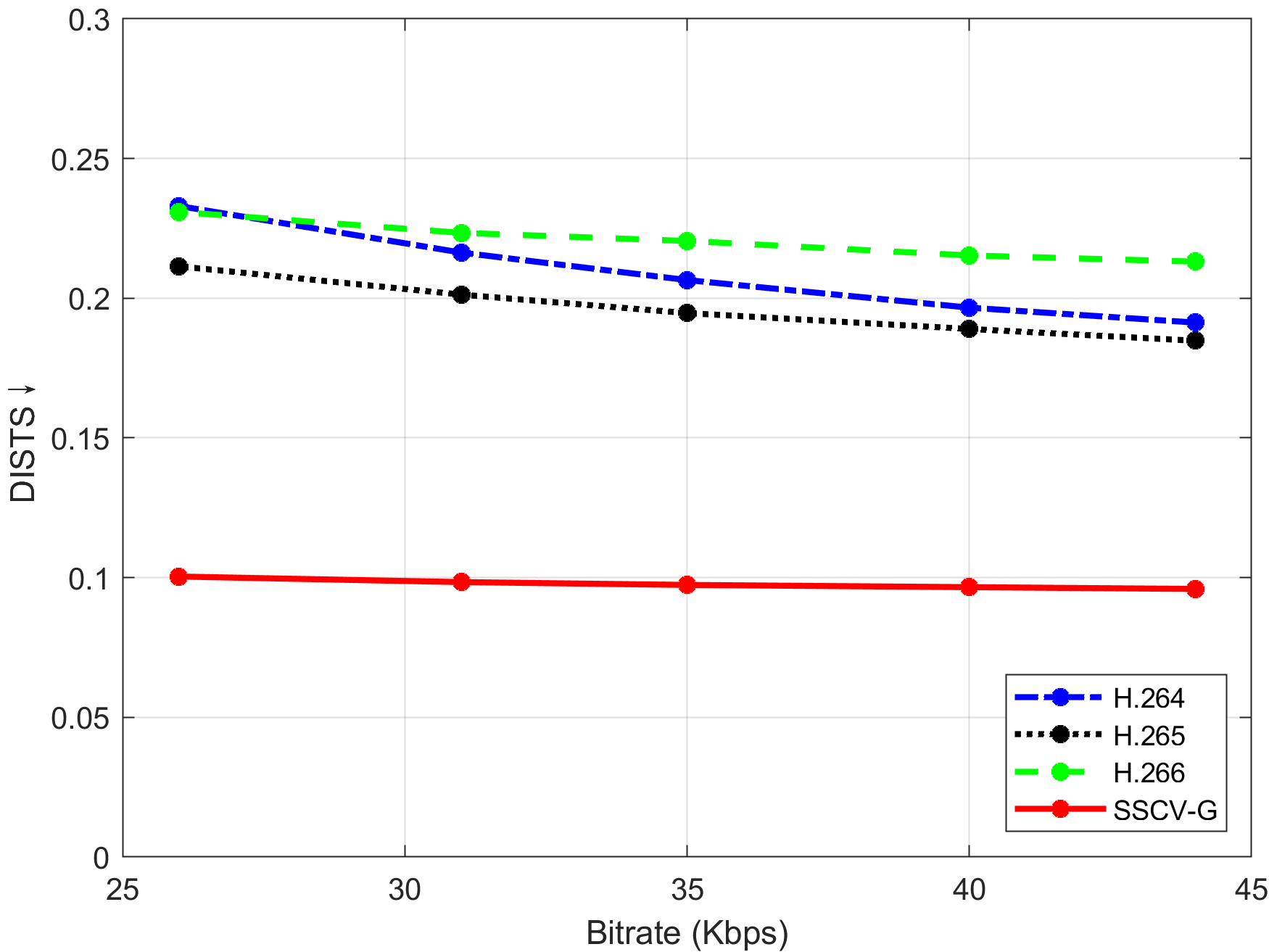}}%
\label{dists}
\captionsetup{justification=raggedright}
\caption{Performance versus different bitrates.}
\label{performance_bitrate}
\end{figure*}

\begin{figure*}[t]
\centering
\subfloat[]{\includegraphics[width=0.24\textwidth]{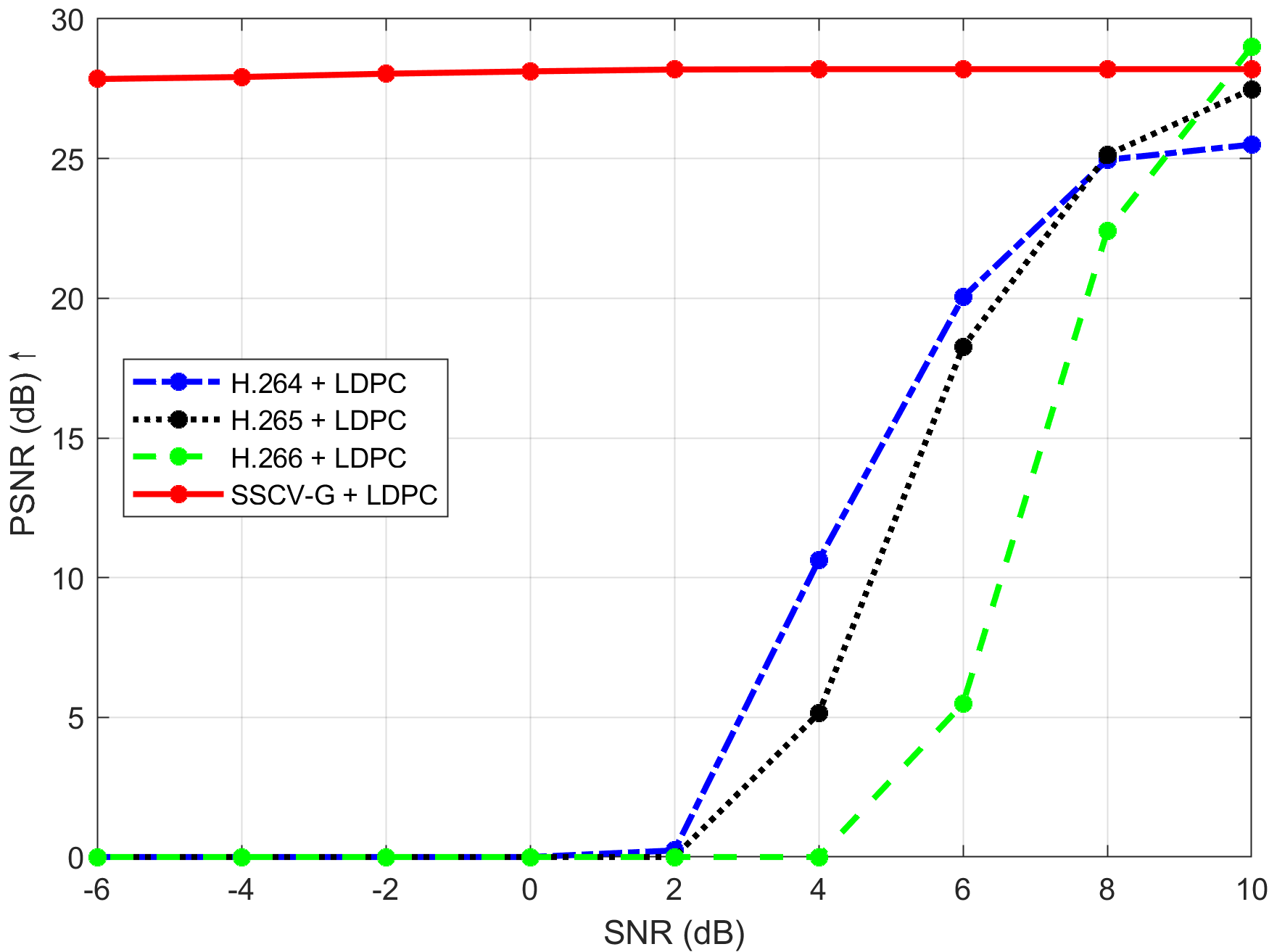}}%
\label{psnr}
\hfil
\subfloat[]{\includegraphics[width=0.24\textwidth]{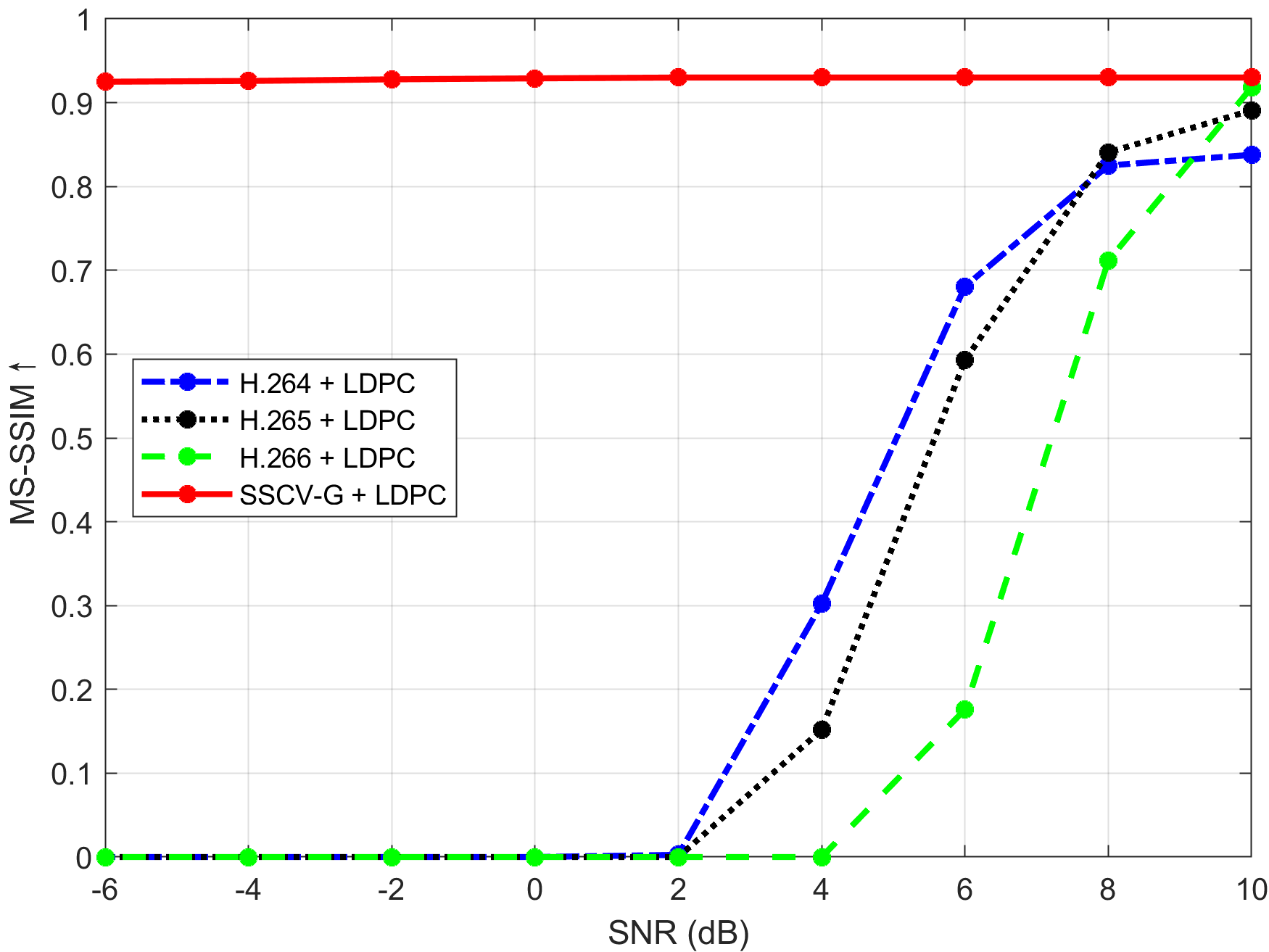}}%
\label{ms-ssim}
\hfil
\subfloat[]{\includegraphics[width=0.24\textwidth]{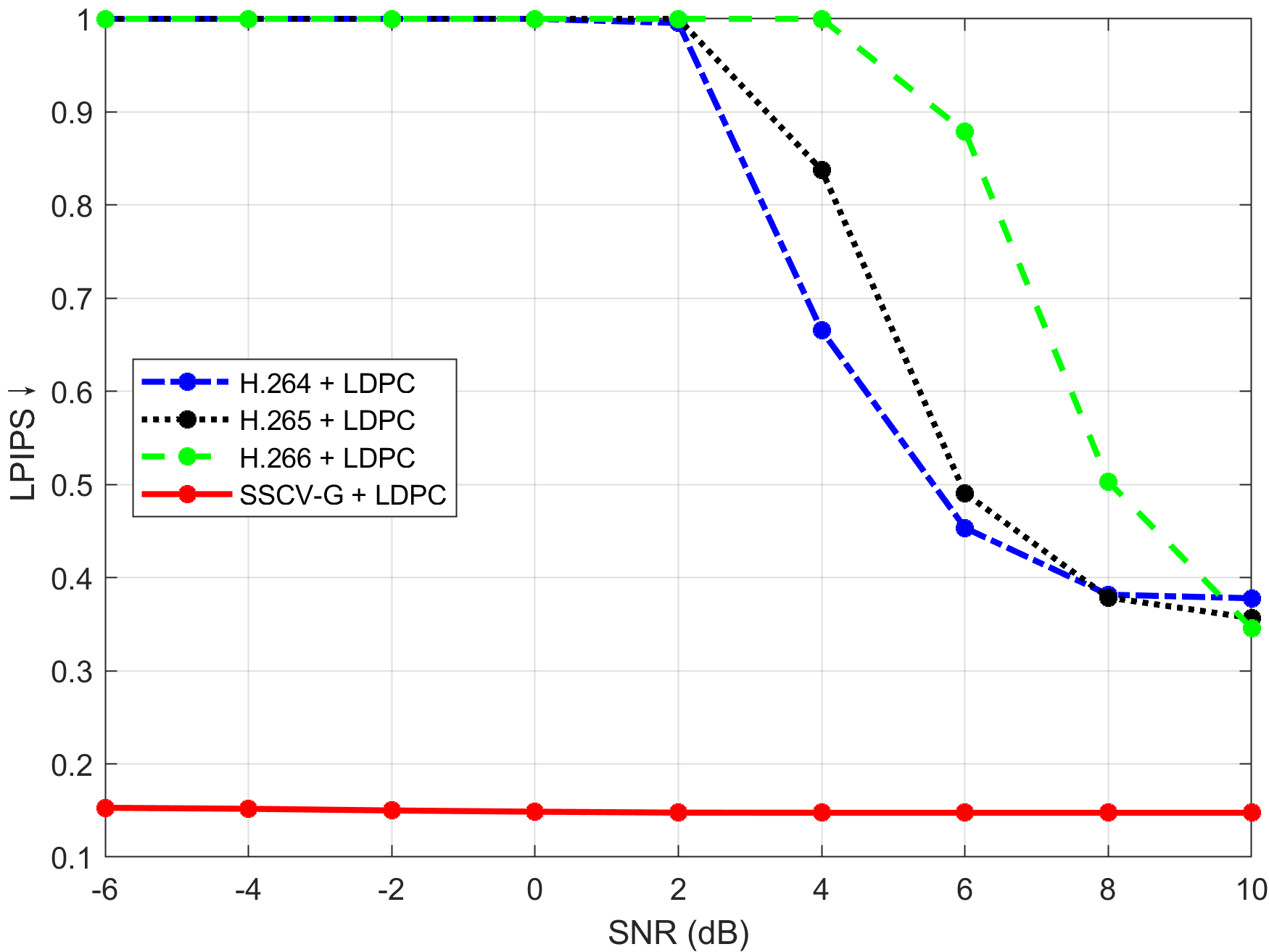}}%
\label{lpips}
\subfloat[]{\includegraphics[width=0.24\textwidth]{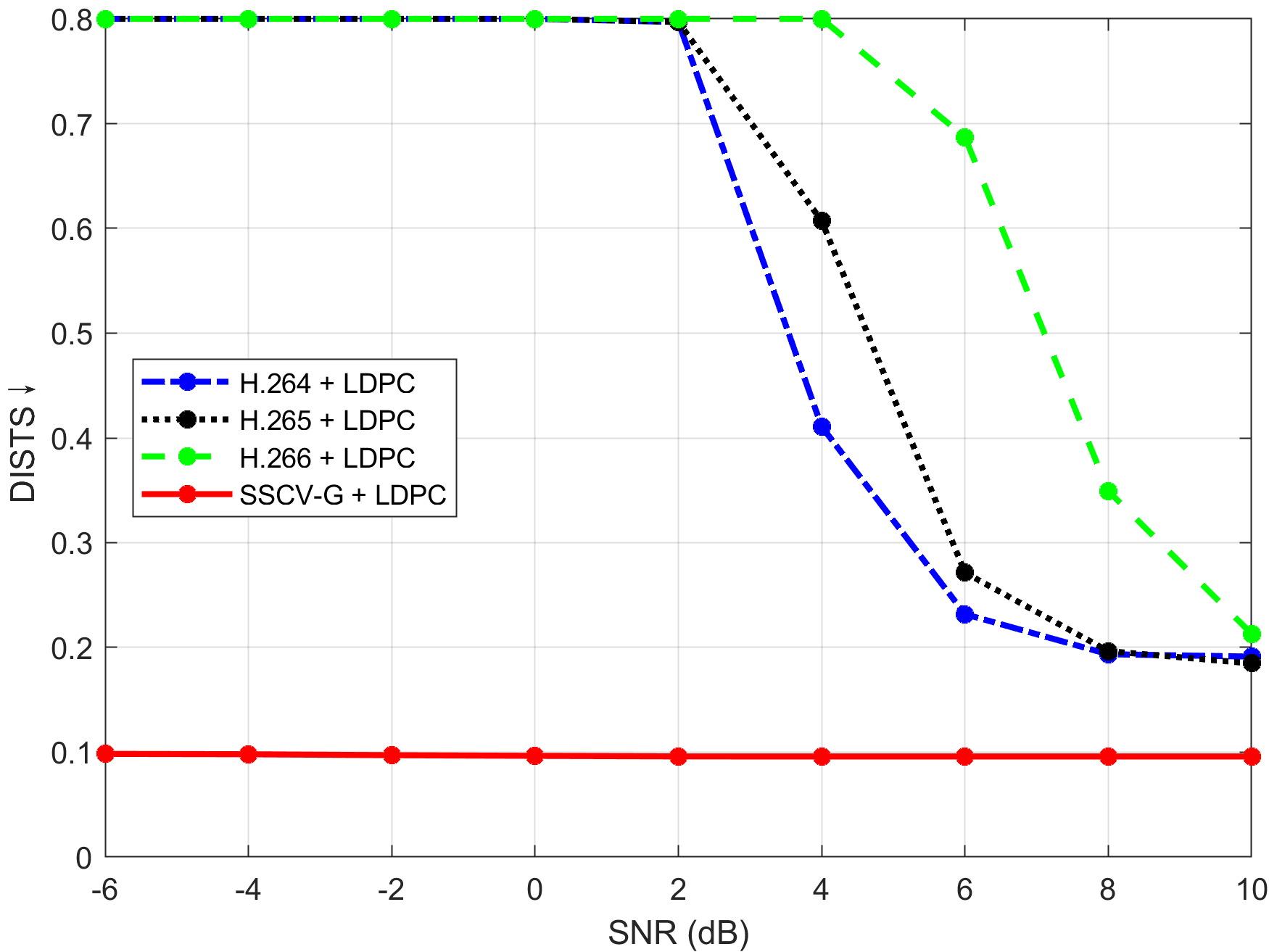}}%
\label{dists}
\captionsetup{justification=raggedright}
\caption{Performance versus different SNRs over AWGN channel.}
\label{performance_snr}
\end{figure*}
\begin{figure*}[t]
    \centering
    \includegraphics[width=0.9\textwidth]{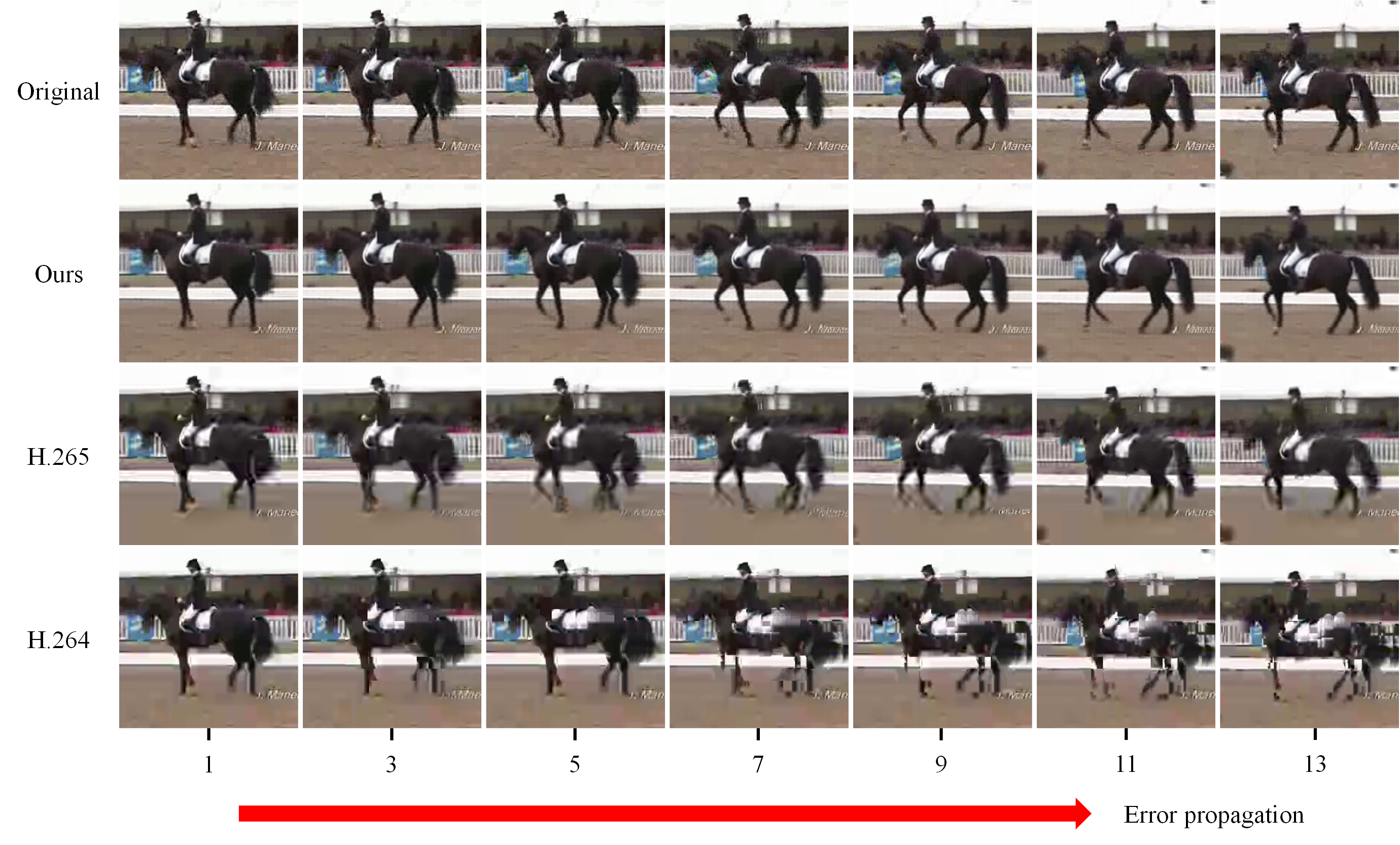}
    \caption{Visualization examples under AWGN channel with the SNR of 8 dB.}
    \captionsetup{justification=raggedright}
    \label{visual_comparision}
\end{figure*}
\section{Experimental Results}
This section will describe the experimental setup and comparison schemes for the proposed video semantic transmission algorithm, SSCV-G, and the resource allocation algorithm MUPPO. Additionally, the performance of the proposed SSCV-G will be demonstrated, and the QoE performance of MUPPO's resource optimization will be explored.
\subsection{Experimental Setup}
\textit{1) Implementation Details:} The UCF-50 dataset is adopted to train the SSCV-G model. This dataset consists of real-world clips from 50 different action categories, covering a wide range of activities including daily routines, sports, and entertainment. To alleviate the computational cost of training, we choose the \textit{HorseRiding} dataset for our experiments. We use all video frames as the dataset for the semantic compression and reconstruction network, and package every $N$ consecutive frames for the loss recovery network. All input video frames are uniformly cropped to 240$\times$240 resolution for consistency during training and evaluation. The test dataset does not overlap with the training dataset, with a test-to-training dataset ratio of approximately 1:9. In the UAV collaborative video streaming scenario, there are four UAVs, the monitored area radius $r_0$ is 100 m, the maximum flight height is 50 m, the minimum flight height is 10 m, the carrier frequency is 2.4 GHz, and the noise power spectral density $\sigma^2$ is -174 dBm/Hz. Additional training parameters for the SSCV-G and MUPPO algorithms are summarized in Table \ref{training_parameters}. 

\textit{2) Comparison Schemes:} All experiments are conducted on a platform equipped with an NVIDIA RTX 4090 GPU and an Intel i9-13900K CPU. To validate the advantages of the proposed SSCV-G algorithm, we use widely implemented H.264 (AVC), H.265 (HEVC), and H.266 (VVC) as video codecs, and low-density parity-check (LDPC) as the channel codec to test the performance of the proposed algorithm in real digital communication systems. The 1/2 rate (500,1000) LDPC code is employed in conjunction with binary phase-shift keying (BPSK) modulation. To validate the advantages of the proposed MUPPO algorithm, we use the dueling double deep Q network (D3QN) algorithm as a benchmark. For the adaptive coding benchmark, we adopt SVC, with 4 enhancement layers, enabling 5 levels of bitrate adaptation to improve bandwidth utilization.

\textit{3) Evaluation Metrics:} Peak signal-to-noise ratio (PSNR) and multi-scale structural similarity index (MS-SSIM) are used to evaluate the performance of SSCV-G in terms of traditional distortion-based metrics, while learned perceptual image patch similarity (LPIPS) and deep image structure and texture similarity (DISTS) are employed to assess perceptual quality. The PSNR is calculated by
\begin{equation}
    \text{PSNR}=10\log_{10} \frac{\text{MAX}^2}{\text{MSE}},
\end{equation}
where $\text{MAX}$ denotes the maximum possible pixel value, and $\text{MSE}$ is the mean square error between the original and reconstructed video frame. The MS-SSIM is computed by
\begin{equation}
    \left[\mathcal{L}_M(\cdot)\right]^{\alpha_M} \cdot \prod_{j=1}^M\left[\mathcal{C}_j(\cdot) \cdot \mathcal{S}_j(\cdot)\right]^{\alpha_j},
\end{equation}
where $M$ denotes the number of resolution levels. $\mathcal{L}_M(\cdot)$ is the luminance similarity at the highest, while $\mathcal{C}_j(\cdot)$ and $\mathcal{S}_j(\cdot)$ represent the contrast and structure similarities at the $j$-th scale, respectively. The exponents $\alpha_M$ and $\alpha_j$ are weighting parameters used to balance the contribution of each component. MS-SSIM extends SSIM by incorporating multi-scale structural comparisons to better capture variations in visual perception. LPIPS evaluates perceptual similarity by computing the distance between deep features extracted from two images using a pre-trained neural network, thereby modeling human visual sensitivity to perceptual differences. DISTS is a composite metric that combines the strengths of SSIM and deep feature representations: it retains SSIM’s sensitivity to structural variations while leveraging deep features to capture fine-grained texture details. Both LPIPS and DISTS use VGG-16 as the backbone for feature extraction.

\subsection{Performance of the  Proposed SSCV-G}
Fig. \ref{performance_bitrate} presents the variation of several performance metrics with respect to bitrate. The proposed algorithm significantly outperforms all the conventional coding schemes in terms of MS-SSIM, LPIPS, and DISTS, and achieves performance comparable to the VVC codec in PSNR. Fig. \ref{performance_snr} shows the impact of SNR on performance across various metrics. While remaining compatible with standard digital communication systems, the proposed method effectively mitigates the "cliff effect" typically observed in traditional source–channel coding schemes. Notably, even under extremely low SNR conditions, the performance of our approach remains stable. This robustness is attributed to multi-frame joint decoding at the receiver, which leverages inter-frame semantic correlations to correct potential semantic errors.

Fig. \ref{visual_comparision} illustrates the superiority of the proposed SSCV-G algorithm. Under 8 dB SNR in AWGN channel, the reconstruction quality achieved by our method significantly outperforms that of state-of-the-art codecs, including H.264 and H.265. Moreover, traditional coding schemes suffer from error propagation due to reference frame corruption caused by channel noise, and subsequent frames decoded based on erroneous references inherit and amplify the errors. In contrast, the proposed algorithm exploits semantic correlations across multiple consecutive frames at the decoder, enabling effective correction of semantic errors introduced during wireless transmission.

\begin{figure}[t]
    \centering
    \includegraphics[width=\linewidth]{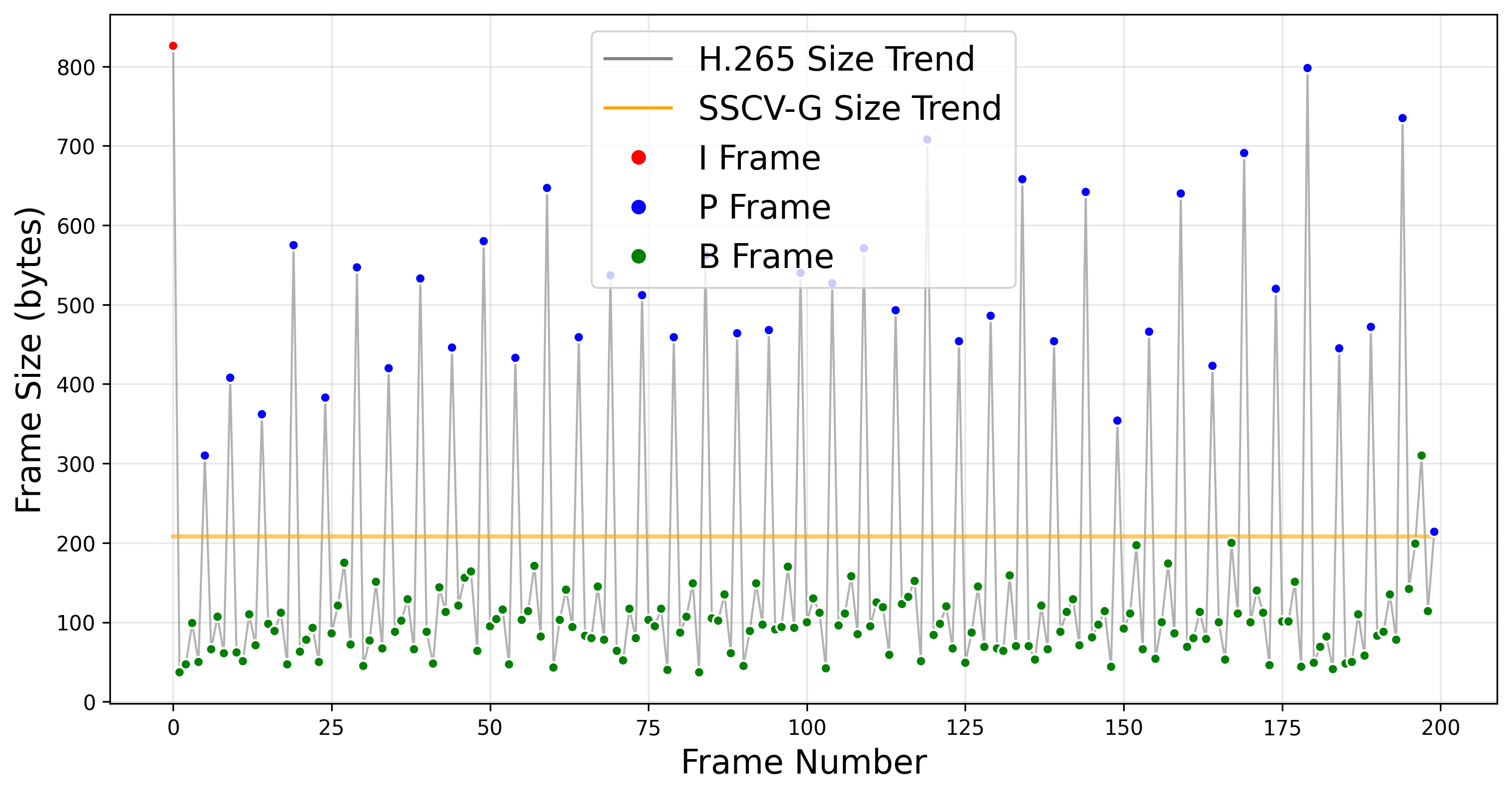}
    \caption{Frame size of the proposed SSCV-G algorithm and H.265. The video is encoded at a bitrate of 50 kbps.}
    \captionsetup{justification=raggedright}
    \label{frame_size}
\end{figure}

Fig. \ref{frame_size} illustrates the variation in frame sizes for the SSCV-G algorithm and H.265 video at a target bitrate of 50 kbps. As shown, there is a significant disparity in bitrate between I frames and both P and B frames. Due to the low bitrate setting, the encoder produces a relatively large GOP, which limits the adaptability of conventional encoding schemes to rapidly changing wireless network conditions. Furthermore, when the average output bitrate approaches the channel capacity, the instantaneous bitrate of I frames may exceed the available bandwidth, resulting in decoding errors or even decoding failure for these frames. Such failures can lead to severe error propagation throughout the video stream.

\begin{figure}
    \centering
    \includegraphics[width=\linewidth]{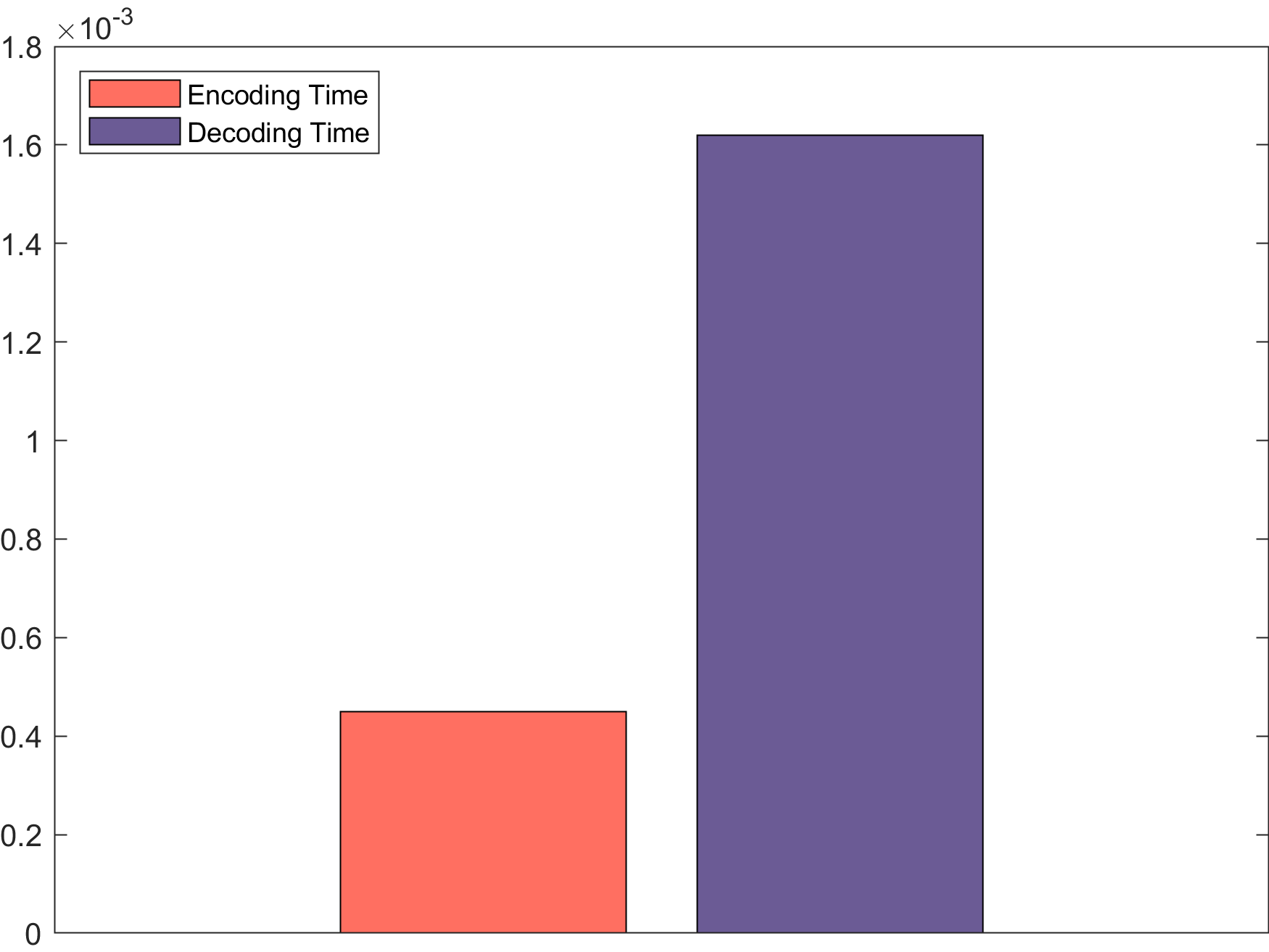}
    \caption{Encoding and decoding latency of the proposed SSCV-G algorithm on the UCF-50 dataset.}
    \captionsetup{justification=raggedright}
    \label{codec_time}
\end{figure}

Fig. \ref{codec_time} presents the encoding and decoding latency of the proposed SSCV-G algorithm on our test platform. The results demonstrate that the proposed method fully supports high-frame-rate real-time video streaming in terms of latency. Although employing more sophisticated encoder architectures or increasing the number of spatio-temporal transformer blocks could further enhance performance, such improvements would inevitably come at the cost of higher encoding and decoding latency.
\subsection{Performance of the  Proposed MUPPO}
\begin{figure}[t]
\centering
\subfloat[Convergence of the proposed MUPPO algorithm.]{\includegraphics[width=\linewidth]{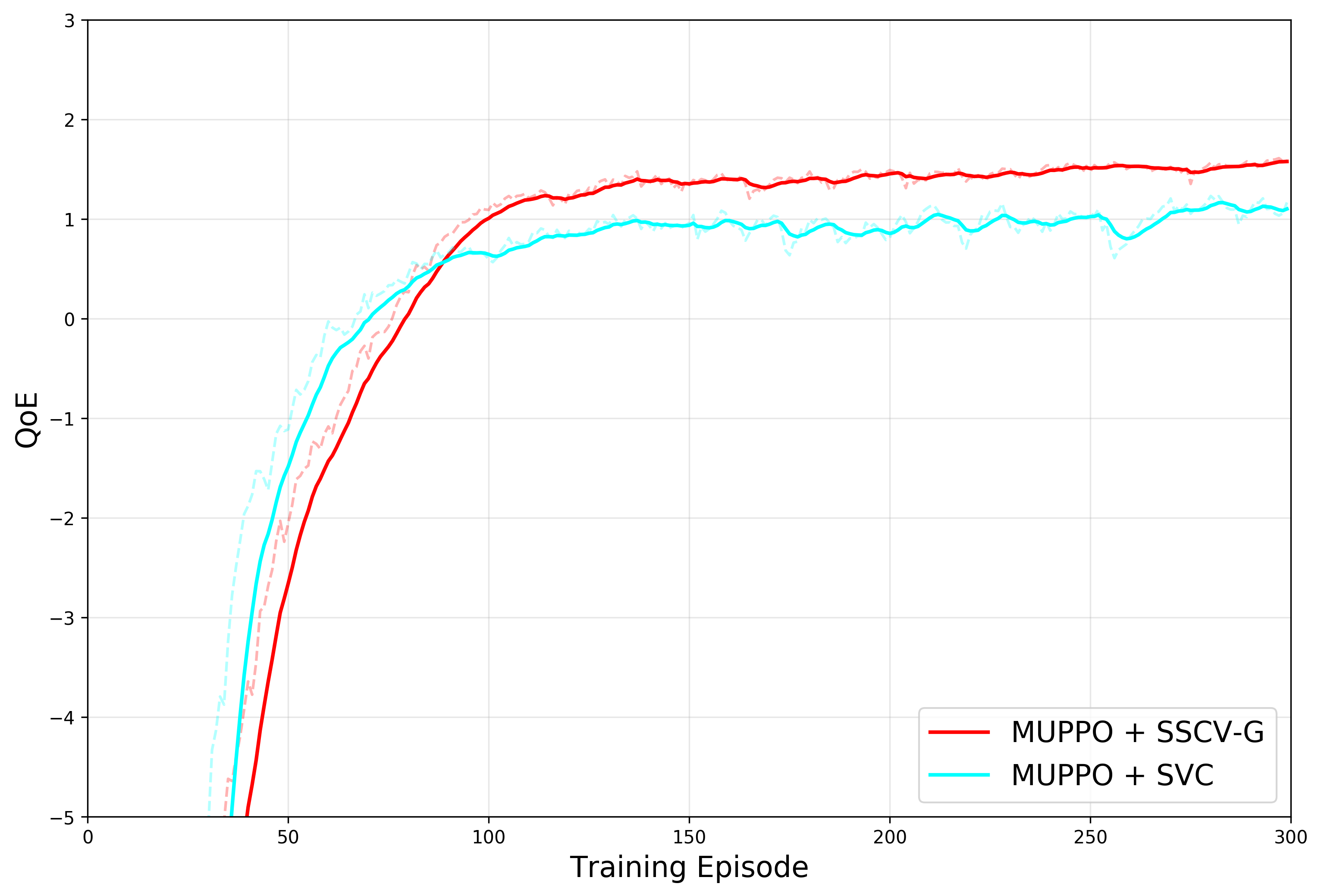}}%
\\
\hfil

\subfloat[Convergence of the D3QN algorithm.]{\includegraphics[width=\linewidth]{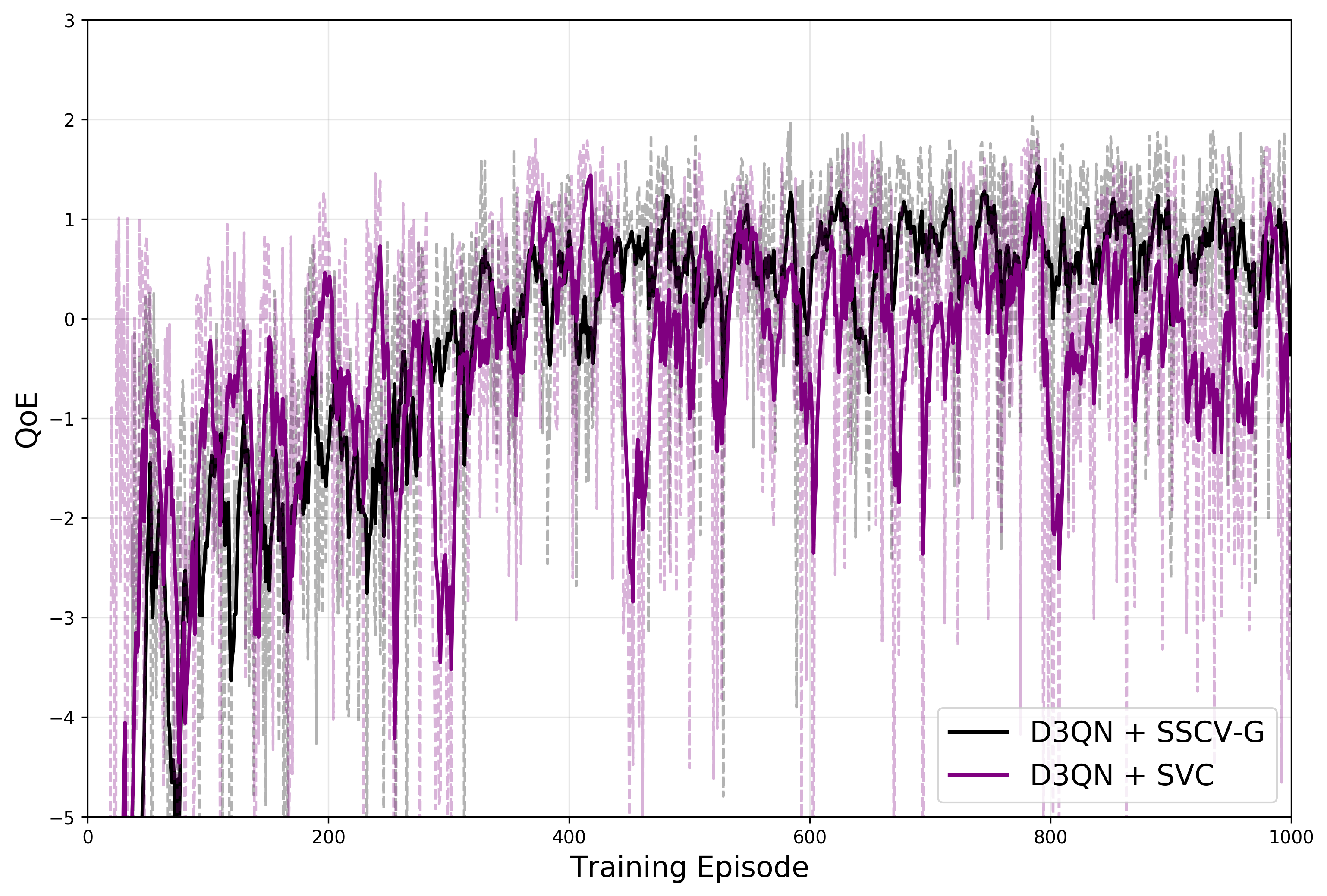}}%
\\
\captionsetup{justification=raggedright}
\caption{Convergence Performance.}
\label{qoe_convergence}
\end{figure}

D3QN is an efficient algorithm in deep reinforcement learning that integrates the enhancements of both Double DQN and Dueling DQN, addressing key limitations of the traditional DQN framework. Doing so improves both convergence speed and policy quality. Therefore, to further evaluate the effectiveness of our proposed SSCV-G algorithm and the MUPPO optimization, we consider four comparative schemes: 1) MUPPO with SSCV-G, 2) D3QN with SSCV-G, 3) MUPPO with SVC, and 4) D3QN with SVC. The convergence performance of these four schemes is illustrated in Fig. \ref{qoe_convergence}.

Compared with D3QN, MUPPO demonstrates faster and more stable convergence, which can be attributed to two main factors. First, we design a multi-threaded environment-interaction mechanism for MUPPO, where each thread independently collects environment and reward data, and all interaction information is aggregated during policy updates. This design significantly mitigates the inefficiency of on-policy algorithms in exploring the environment. Second, MUPPO directly optimizes the cumulative return using policy gradient methods and employs a clipping function to constrain policy updates, ensuring smoother updates. In contrast, D3QN relies on temporal-difference learning, which typically requires a large number of samples to adequately cover the state space. This can result in unstable learning when exploration is insufficient or when rewards are sparse, ultimately reducing learning efficiency.

\begin{figure}[t]
    \centering
    \includegraphics[width=\linewidth]{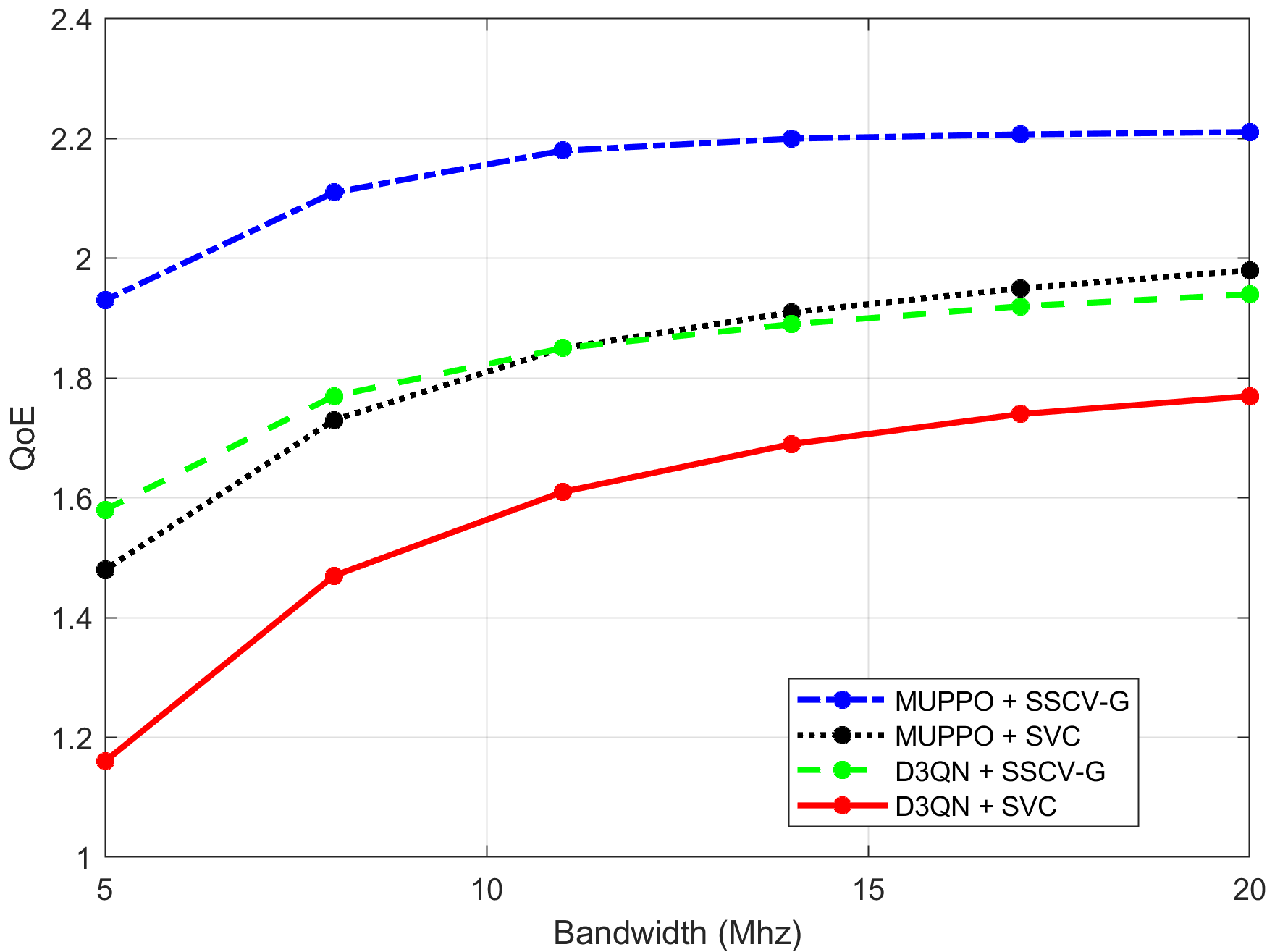}
    \caption{QoE performance versus different bandwidth.}
    \captionsetup{justification=raggedright}
    \label{qoe_bw}
\end{figure}

\begin{figure}[t]
    \centering
    \includegraphics[width=\linewidth]{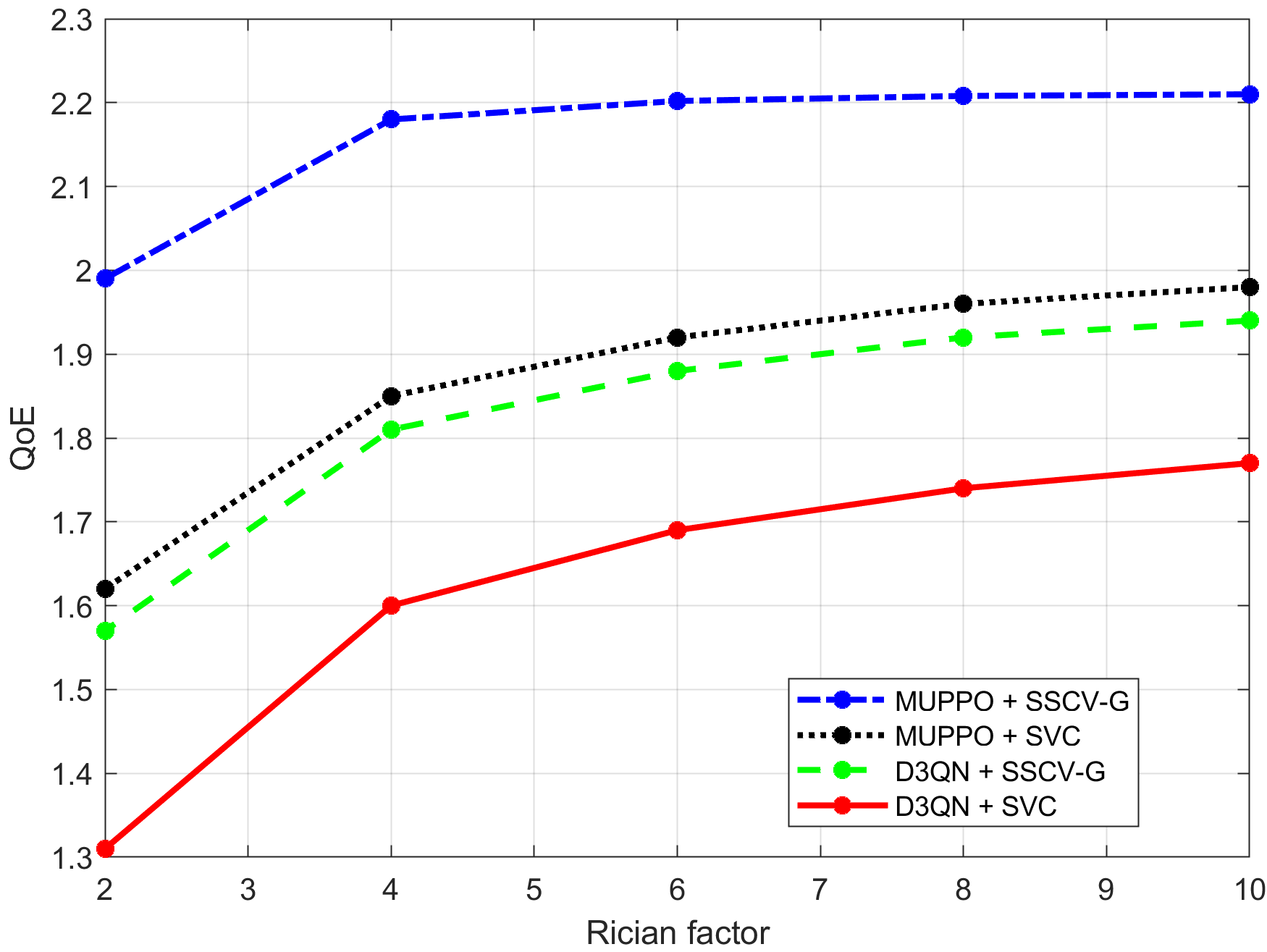}
    \caption{QoE performance versus different Rician factor.}
    \captionsetup{justification=raggedright}
    \label{qoe_rician}
\end{figure}

Figs. \ref{qoe_bw} and \ref{qoe_rician} illustrate the variation of QoE with respect to bandwidth and the Rician factor, respectively. Each test experiment is conducted over 100 independent runs, and the mean performance is reported. As shown in the figures, the proposed MUPPO + SSCV-G algorithm consistently outperforms all the benchmark methods. In particular, our approach demonstrates strong robustness to bandwidth fluctuations. When the bandwidth drops to 5 MHz, the QoE of our method decreases by only 12.7\%, whereas the competing methods exhibit declines exceeding 18.6\%. When the Rician factor drops to 2, the QoE of our method decreases by only 10.0\%, while the competing benchmarks exhibit declines exceeding 18.2\%. This performance gain is attributed to the precise rate control and packet loss resilience of the SSCV-G algorithm, as well as the robustness of the MUPPO scheme.

\section{Conclusion}
In this paper, we propose a novel semantic self-correcting video encoding and transmission framework with ultra-fine-grained bitrate adaptability (SCCV-G). By sharing a semantic codebook between transmitter and receiver, video frames are represented as semantic indices, enabling highly efficient compression. The transmitter selectively discards semantic indices based on real-time bandwidth constraints, allowing precise bitrate control via flexible discard ratios. To recover missing semantics, the receiver employs a multi-frame joint decoding strategy that exploits temporal semantic correlations. To further enhance the performance of SSCV-G in multi-user and dynamic UAV environments, we integrate a multi-user proximal policy optimization (MUPPO) algorithm into the framework. MUPPO jointly optimizes communication resource allocation, semantic bitrate selection, and UAV trajectory planning, serving as a high-level control module that adaptively guides SSCV-G toward QoE-aware decisions. Extensive evaluations demonstrate that our joint SSCV-G and MUPPO framework significantly outperforms state-of-the-art video coding schemes in compression efficiency, adaptability to bandwidth variation, and robustness to transmission noise. Notably, our method improves overall user QoE by 13.92\% compared to competitive baselines.
\bibliographystyle{IEEEtran}%
\bibliography{ref}  %

\end{document}